\documentclass[useAMS,usenatbib]{mn2e}
\usepackage{amssymb, hyperref, rotating, multirow, threeparttable, mathptmx, fixltx2e, graphicx}
\def\aj{AJ} \def\apj{ApJ} \def\apjs{ApJS} \def\mnras{MNRAS}
\def\apjl{ApJ} \def\nat{Nature} \def\araa{ARA\&A} \def\aap{A\&A}
   
\let\ga=\gtrsim \let\la=\lesssim
\def\Msun{~M$_{\sun}$}     

\title[Hot gas quenching]
{Hot gas in massive halos drives both mass quenching and environment quenching}

\author[Gabor \& Dav\'e]{
J. M. Gabor,$^{1}$\thanks{Email:jared.gabor@cea.fr}
R. Dav\'e$^{2,3,4}$
\\ $^{1}$CEA-Saclay, IRFU, SAp, F-91191 Gif-sur-Yvette, France  \\
 $^{2}$University of the Western Cape, Bellville, Cape Town 7535, South Africa \\
 $^{3}$South African Astronomical Observatories, Observatory, Cape Town 7925, South Africa \\
 $^{4}$African Institute for Mathematical Sciences, Muizenberg, Cape Town 7945, South Africa
}

\begin{document}


\pagerange{\pageref{firstpage}--\pageref{lastpage}} \pubyear{2013}

\maketitle
\label{firstpage}

\begin{abstract}
Observed galaxies with high stellar masses or in dense environments
have low specific star formation rates, i.e. they are quenched.  Based
on cosmological hydrodynamic simulations that include a prescription
where quenching occurs in regions dominated by hot ($>10^{5.4}$~K) gas, we
argue that this hot gas quenching in halos $>10^{12}$\Msun drives both
mass quenching (i.e. central quenching) and environment quenching
(i.e. satellite quenching).  These simulations
reproduce a broad range of locally observed trends among quenching,
halo mass, stellar mass, environment, and distance to halo center.
Mass quenching is independent of environment because $\sim
10^{12}-10^{13}$\Msun ``mass quenching halos'' inhabit a large range
of environments.  On the other hand, environment quenching is
independent of stellar mass because galaxies of all stellar masses may
live in dense environments as satellites of groups and clusters.  As
in observations, the quenched fraction of satellites increases with
halo mass and decreases with distance to the center of the group or
cluster.  We investigate pre-processing in group halos, ejected former
satellites, and hot gas that extends beyond the virial radius.  The
agreement of our model with key observational trends suggests that hot
gas in massive halos plays a leading role in quenching low-redshift
galaxies.
\end{abstract}

\section{Introduction} \label{sec.intro}

It has long been recognized that the most massive galaxies as well as
galaxies in dense regions have not formed substantial new stars for
many Gyr.  As a result they lie on a tight locus in color-magnitude
space known as the red sequence.  The mechanisms that halt star
formation and cause the appearance of the red sequence of galaxies,
often referred to as quenching, remain poorly understood.
Observations imply that the red sequence noted at $z=0$
\citep[e.g.][]{strateva01} is already well-established by $z=1$
\citep[e.g.][]{bell04, arnouts07, faber07} and even earlier at
$z\sim2$ \citep{williams09, brammer09, brammer11, whitaker13} to
$z\sim 4$ \citep{ilbert13,muzzin13, straatman13}.  Deep surveys enable
careful accounting of the growth of the red sequence over cosmic time.
Combined with the present-day age of the stellar population in red
sequence galaxies, this suggests that they must have built up mass
through star-formation at an early epoch, and then some process(es)
quenched their star formation.

A common theoretical explanation for quenching is that massive halos
above $\sim10^{12}$\Msun form hot gas coronae via the virial shock
\citep{birnboim03, keres05}, and this hot gas fails to cool, thus
starving galaxies of fuel for star-formation \citep{dekel06,
cattaneo06, birnboim07}.  The virial shock heats any gas inflows
from the inter-galactic medium (IGM), preventing that inflowing gas
from directly feeding the galaxies.  In many circumstances the hot
gas is expected to eventually cool in a so-called cooling
flow~\citep[e.g.][]{fabian84}, so an additional heating source is required
to prevent such cooling and resulting star-formation \citep[e.g.]{croton06,
bower06, somerville08}.  This general class of quenching model is
sometimes called halo quenching, owing to the dependence on halos
above $10^{12}$\Msun; or radio mode AGN feedback \citep{croton06},
owing to the invocation of radio AGN as the heating source; or
maintenance mode feedback, because an AGN only needs to maintain
the high temperature of the halo gas rather than heat it from cold
to hot.  We prefer ``hot halo'' quenching because we will argue
that the hot gaseous halo is the key ingredient.  

Depending on their density structure, some hot halos may naturally
have cooling times longer than the Hubble time, so no feedback
effects are needed to quench star-formation~\citep{keres09_coldmode}.
Furthermore, unresolved gravitational heating may provide sufficient
thermal energy to maintain hot halos without the need for black
hole feedback \citep{dekel08, johansson09, birnboim11, feldmann11}.
The details of heating and cooling of hot gas in massive halos are
not fully understood~\citep{delucia10}, but models where hot gaseous
halos are assumed to stay hot can successfully produce a realistic
red sequence \citep{gabor11}.

Observed correlations between quenched galaxies and stellar mass,
morphology, and environment offer clues to their formation mechanisms
(e.g. \citealt{blanton09} for a review for local galaxies;
\citealt{elbaz07}, \citealt{cooper07}, and \citealt{quadri12} for
environmental effects at $z\approx 1-2$).  Recent work has revealed
trends suggesting two distinct quenching modes -- ``mass quenching''
and ``environment quenching'' \citep{peng10}.  These appear distinct
because galaxies are increasingly likely to be quenched at high
stellar masses, independent of environment (mass quenching), and
galaxies are increasingly likely to be quenched in dense environments,
independent of stellar mass (environment quenching).  \citet{peng12}
argue that environment quenching applies mainly to satellite galaxies,
and appears independent of halo mass.  Based on these local trends, they
suggest that a quenching model based on halo mass is problematic.
\citet{knobel13} suggest that these trends persist out to $z\sim1$.
\citet{woo13} reframed satellite quenching by considering distance to
the halo center, and argue that halo quenching models could explain
the results.

In this work we use cosmological hydrodynamic simulations to show
how quenching associated with hot halo gas drives the observed trends.
Specifically, both mass quenching and environment quenching emerge
from the same simple quenching model, where galaxies in hot halos
are quenched.  Because the presence of hot gas is closely tied to
halo masses $>10^{12}$\Msun~\citep[e.g.][]{birnboim03,keres05,gabor10}, this
implies that ``halo quenching'' models remain viable.

We describe our simulations in \S\ref{sec.sims}, including details of
our quenching model in \S\ref{sec.quench_model}.   Our simulation
  methodology is based on that in \citet{gabor12}, but we use a larger
  simulation volume.  With this larger simulation we extend our
  previous analysis of the growth of the red sequence to explore the
  relationships between hot gas and environment.  Inspired by
observations, we illustrate various trends among quenching, halo mass,
stellar mass, and environment, in \S\ref{sec.trends}.  These trends
include the red fraction as a function of environment and stellar
mass, the stellar mass -- halo mass relation for centrals and
satellites, and the invariance of star-forming galaxy stellar mass
functions with halo mass.  In \S\ref{sec.discussion} we lay out a
quenching narrative for galaxies in different halos, and we conclude
with \S\ref{sec.conclusion}.

\section{Simulations}
\label{sec.sims}
\begin{figure}
\includegraphics[width=79mm]{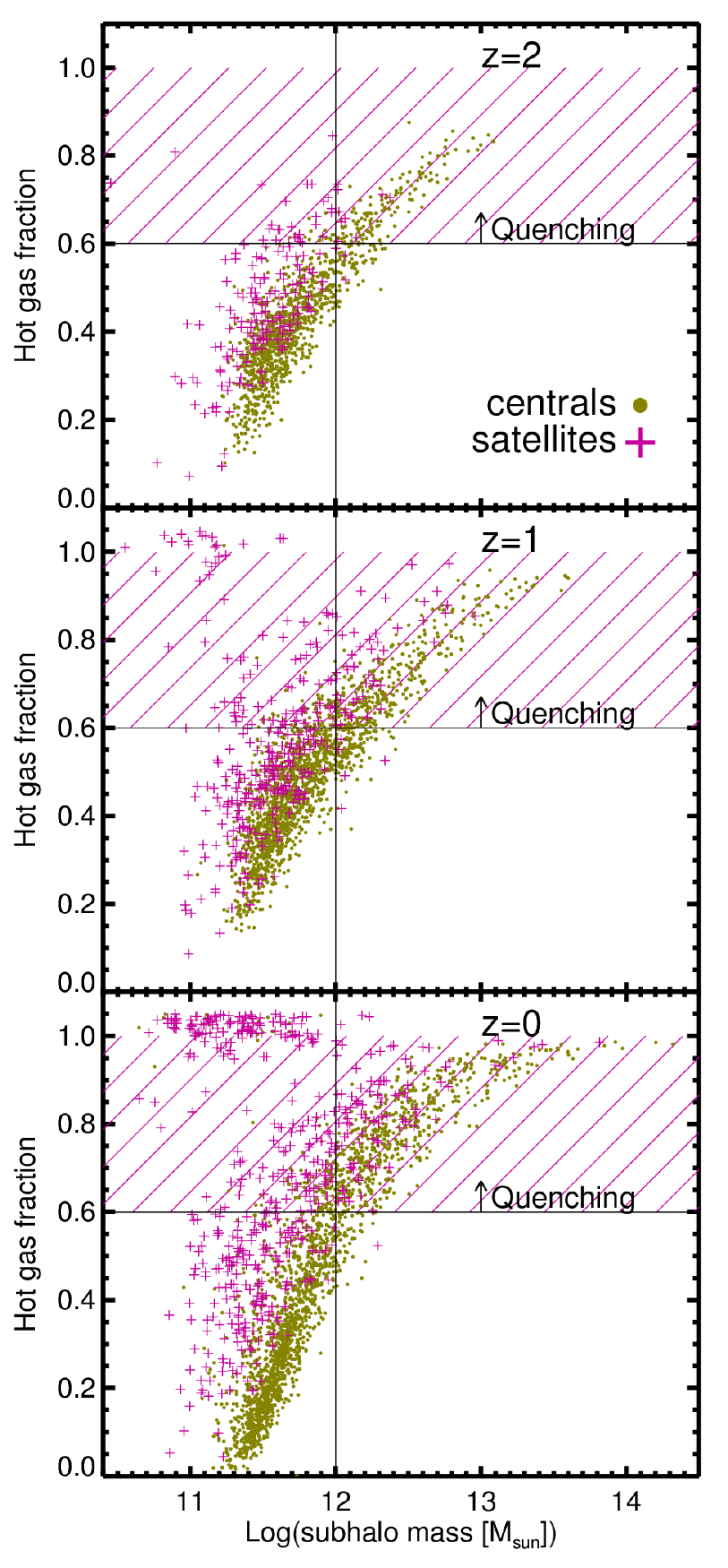}
\caption{Hot gas fraction $f_{\rm hot}$ vs. subhalo mass at $z=2,1,$
  and $0$ (top to bottom).  Thin black lines mark $M_{\rm
    halo}=10^{12}$~\Msun and $f_{\rm hot}$ for visual reference.  For
  centrals (green circles), hot gas fraction correlates well with halo
  mass, and halos above $\sim10^{12}$~\Msun are dominated by hot gas.
  Many satellite galaxies (purple crosses) are dominated by hot gas
  despite low subhalo masses -- these generally have massive parent
  halos that contain hot gas (we add a small amount of scatter to
  satellites with $f_{\rm hot}=1$ for clarity).  The trends show
  little variation with redshift.  In our quenching model, we prevent
  gas accretion onto galaxies with $f_{\rm hot}>60$\%., as indicated
  by diagonal hatching.}
\label{fig.hotfrac_vs_mass}
\end{figure}
%
%
%
\begin{figure*}
\includegraphics[width=168mm]{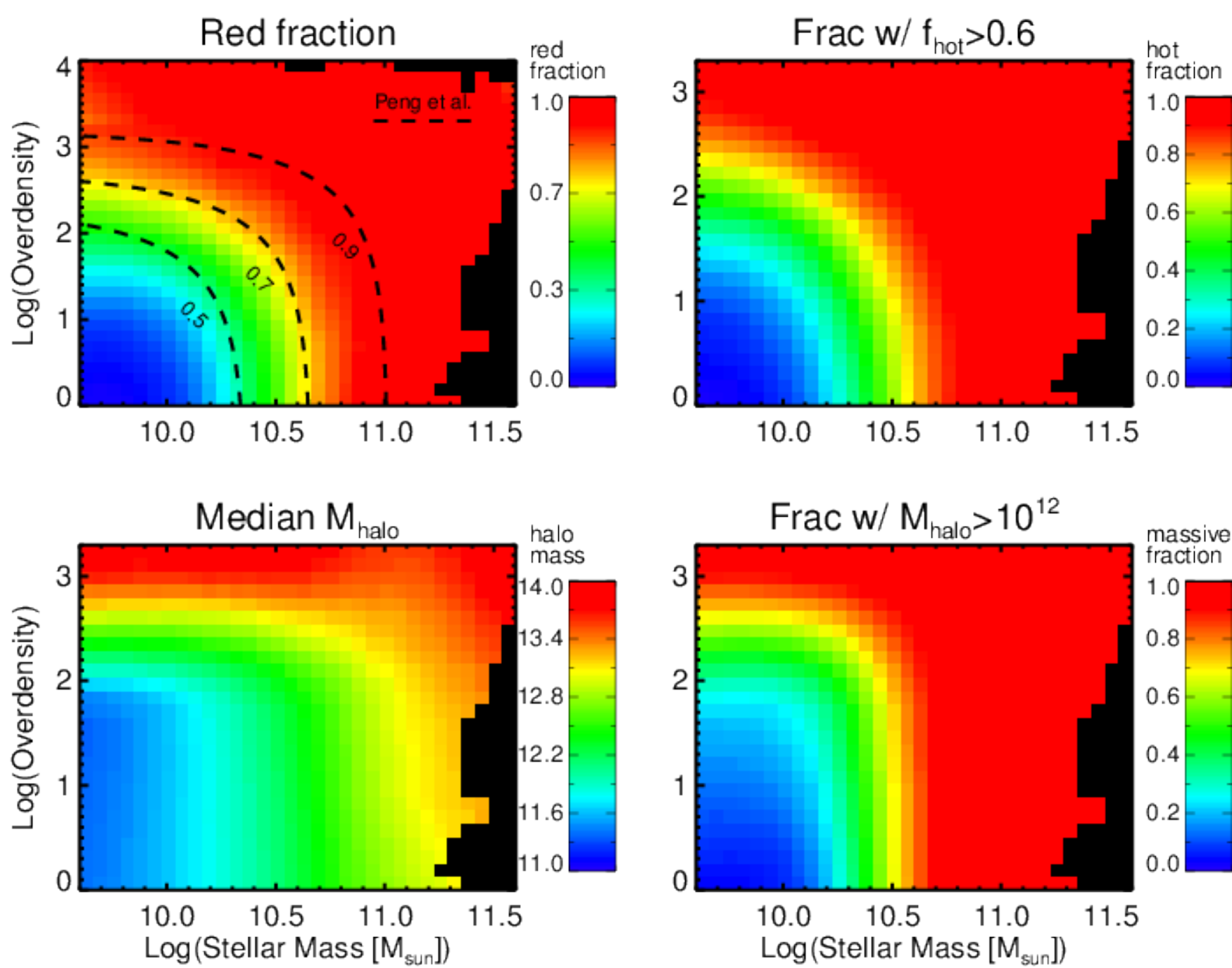}
\caption{The appearance of mass quenching and satellite quenching in
  our simulations: as a function of 5th-nearest neighbor overdensity
  (see \S \ref{sec.analysis}) and stellar mass, we plot (in color) the
  fraction of galaxies that are quenched ({\bf top left}), the
  fraction of galaxies whose {\it subhalo} is dominated by hot gas
  ({\bf top right}), the median parent halo mass ({\bf bottom left}),
  and the the fraction of galaxies with a parent halo mass above
  $10^{12}$\Msun ({\bf bottom right}). The boxy shape of the contours
  in observations of the red fraction (cf. dashed lines in top left
  panel) led \citet{peng10} to conclude that mass quenching and
  environment quenching are distinct processes.  This figure shows
  that, in our hot gas quenching model, the boxy pattern appears with
  parent halo mass as well.  Mass quenching and environment quenching
  are two manifestations of quenching in hot, massive halos.}
\label{fig.2dhists}
\end{figure*}
%
\begin{figure}
\includegraphics[width=84mm]{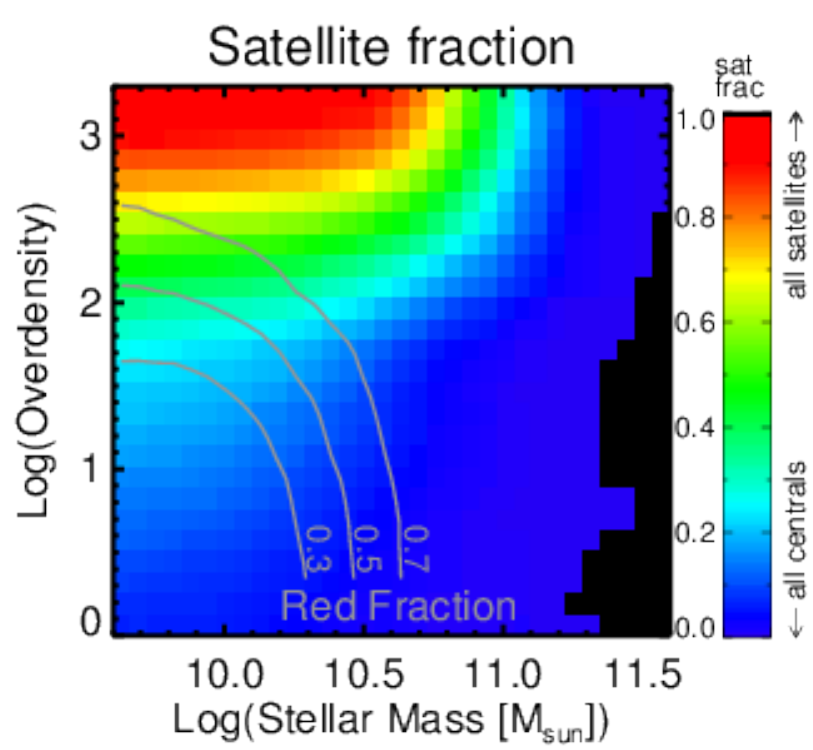}
\caption{Satellite fraction (colors) as a function of environment and
  stellar mass.  Satellite fraction is calculated in each pixel $i$
  (i.e. each 2-dimensional bin of overdensity and stellar mass) as
  $f_{{\rm sat},i} = N_{{\rm sat},i}/N_{{\rm tot},i}$, where $N_{{\rm
      sat},i}$ is the number of satellite galaxies falling in pixel
  $i$ and $N_{{\rm tot},i}$ is the total number of galaxies (centrals
  plus satellites) falling in pixel $i$.  Gray lines show contours of
  red (i.e. quenched) fraction from Figure \ref{fig.2dhists}.  This shows that
  environment quenching is mainly associated with satellite galaxies,
  while mass quenching is associated with centrals.}
\label{fig.fsat}
\end{figure}

The basic methodology of the simulations used here is identical to that in
\citet{gabor12}, albeit with larger volumes.
We summarise the simulations here.

We use a modified version of the publicly-available Smoothed
Particle Hydrodynamics code \textsc{gadget-2} \citep{springel05} to solve
N-body gravitational dynamics and the equations of hydrodynamics
starting from cosmological initial conditions.  We include a model for
gas cooling and photoionization from a meta-galactic UV background
\citet{sutherland93, haardt01}.  We track chemical evolution of gas
and stars through four metal species and our cooling function
incorporates metal-line cooling appropriate for each gas particle's
metallicity \citep{oppenheimer08}.

Gas that collapses into structures denser than a threshold of
0.13~cm$^{-3}$ is governed by the 2-phase star-formation model of
\citep{springel03}, based on \citep{mckee77}.  This
single-parameter model is tuned to match the Kennicutt star-formation
relation \citep{kennicutt98}.  Star-forming gas particles self-enrich
with metals from supernovae, and stochastically spawn collisionless
star particles as well as ejecting galactic winds.  During each time step, each
star-forming gas particle spawns a new star particle with a
probability consistent with its star-formation rate.  The typical mass
of the star particles thus spawned is roughly half the gas particle
mass, so that gas particles generally may spawn two star particles.

After formation, star particles lose mass and metals due to AGB
feedback and Type Ia supernovae.  We calculate evolved stellar mass
loss using stellar population models \citep{bc03}, and spread the lost
mass to nearby gas particles.  Type Ia SN rates are taken from
\citep{scannapieco05}, and these inject both mass and energy into the
surrounding gas.

Star-forming gas particles may also be launched in a galactic wind.
We assume the wind mass outflow rate, $\dot{M}_{\rm wind}$, is
proportional to the star-formation rate divided by the galaxy circular
velocity.  In order to calculate circular velocities, we use an
on-the-fly friends-of-friends (FoF) galaxy-finder to estimate galaxy masses.
During each time step, each star-forming gas particle has a
probability of being launched in a wind that is consistent with
$\dot{M}_{\rm wind}$.  New wind particles are given a kick velocity
$v_{\rm wind}$ proportional to the galaxy circular velocity, and they
are decoupled from the normal hydrodynamics until the local density
has dropped to one-tenth the star-formation threshold density, for up
a time up to 20~kpc/$v_{\rm wind}$.  The scalings of $\dot{M}_{\rm wind}$ and $v_{\rm wind}$
with the SFR and circular velocity are motivated by the theory of
momentum-conserving winds \citep{murray05, murray11}, which in turn is
motivated by local observations of starburst galaxies \citep{rupke05,
martin05}. These wind scalings lead to star-forming galaxies and
inter-galactic medium properties that match a wide variety of
observational constraints \citep{oppenheimer06, finlator06, dave07,
finlator07, finlator08, dave08, oppenheimer08, oppenheimer09_OVI,
oppenheimer09_metal_absorption, dave11_gasfrac, dave11_sfr,
oppenheimer12}.

Simulations using our favoured wind model reproduce the observed number
densities of low-mass star-forming galaxies, but they generally
over-produce massive galaxies.  In part to remedy this problem, we
developed a quenching model for our simulations, as discussed next.

\subsection{Quenching in hot halos}
\label{sec.quench_model}
We quench star-formation by adding thermal energy to hot halos in our
simulations.  Hydrodynamic calculations robustly predict the formation
of a stable ``virial shock'' in dark matter halos (only) above a mass
$\sim10^{12}$\Msun \citep[e.g.][]{birnboim03, keres05, gabor10}.  This virial shock heats the gas in the halo from
typical IGM temperatures of $10^4$~K to $>10^5$~K.  While the hot gas
overall has long cooling times, in the central regions it typically cools 
and fuels star formation in simulations.  In Figure \ref{fig.hotfrac_vs_mass} 
we show the hot gas fraction as a function of subhalo mass in one of our
simulations \emph{without} quenching.  The hot gas fraction for
central galaxies rapidly increases around halo masses of
$10^{12}$\Msun.  The relationship is fairly tight, but at a fixed hot
fraction there is scatter of at least a tenth of a decade in subhalo
mass.  Satellite galaxies often live in the hot gas environment of
their more massive host halos, so that their local hot gas fraction is
decoupled from their subhalo mass.

Additional heating sources such as radio AGN \citep{croton06,
somerville08}, cosmic rays \citep{mathews09}, or unresolved
gravitational heating \citep{dekel08, johansson09, birnboim11} may act
to heat the gas and prevent it from cooling, thus starving galaxies in
the centers of hot halos of fuel for star-formation.  Other processes
such as ram pressure stripping \citep{gunn72} may help quench satellite
galaxies.  Our quenching model is designed to emulate this heating in hot halos,
while remaining agnostic regarding its physical origin.  Our basic
approach is to identify galaxies living in hot halos, then
continuously heat the nearby circum-galactic medium around those galaxies to
prevent it from cooling and forming stars.

We identify hot halos on-the-fly by exploiting the FoF galaxy finder.
The FoF galaxies generally include the stars and star-forming gas as
well as a small fraction of the immediate circum-galactic
(non-star-forming) gas.  For each identified FoF galaxy, we estimate
the dark matter subhalo mass and virial radius by scaling the mass
from the FoF algorithm.  We count up all gas particles within that
virial radius, and calculate the fraction that is hotter than
$10^{5.4}$~K.  We take this value from \citet[cf.][]{keres05},
  who found that the gas flowing into simulated galaxies has a
  temperature distribution with a local minimum at this critical
  temperature, corresponding to a local minimum in the cooling curve
  that separates free-free emission from helium cooling.  If the hot
gas fraction is above 0.6, we consider the galaxy to be embedded in a
hot halo.  We initially chose the critical hot fraction of 0.6 to
match the hot fraction in the average $10^{12}$~\Msun halo, as in
Figure \ref{fig.hotfrac_vs_mass}, and we have not tuned it to match
any observables.  Experiments with higher critical hot fractions
  (e.g. 0.7) show that these allow the formation of more massive
  galaxies, but too few low and moderate-mass galaxies are quenched.
In galaxies embedded in a hot halo, we heat the circum-galactic gas to
the virial temperature.  We do not add heat to the star-forming gas,
and we do not add heat throughout the halo -- we only heat gas
identified by the FoF galaxy finder.  We heat these gas particles at
every time step, using arbitrary amounts of thermal energy to keep
them at the virial temperature.

Recent cosmological simulations using the moving mesh code
\textsc{Arepo}~\citep[cf.][]{springel10} suggest that the nature of
galaxy fueling depends on hydrodynamic method.  At late times,
  moving-mesh simulations tend to show more cooling in massive halos
  than do SPH simulations~\citep[][]{keres12, vogelsberger12},
  probably owing to inaccuracies in standard SPH
  formulations~\citep{hayward14}. Contrary to previous results, the
majority of the gas fueling galaxies in \textsc{Arepo} simulations
passes through a hot phase \citep{nelson13}.  The full implications of
this difference have only begun to be explored \citep{vogelsberger13,
  torrey14}, but the appearance of stable hot halos seems to be mostly
unchanged in these simulations.  In \citet{nelson13}, hot halos
develop at halo masses $<10^{12}$\Msun in both moving mesh and
\textsc{Gadget} simulations, but this is lower than in our simulations
partly because they do not include the effects of metal line
cooling~\citep[and hence are more in line with][]{keres05}.  {We
  therefore suspect that uncertainties related to the hydrodynamic
  method will have only a minor impact on our simple quenching model.
  Although the different methods lead to different temperature and
  density distributions of gas within halos \citep{keres12}, they
  apparently do not drastically change \emph{which} halos are hot.
  Our quenching model triggers based on the latter.  Ongoing tests
  with revised formulations of SPH which improve its accuracy
  \citep{saitoh13, hopkins13_sph} should address this uncertainty more
  directly.}

By heating the circumgalactic gas around galaxies living in hot halos,
our quenching model effectively starves or strangles galaxies by
cutting off the fuel for star-formation.  In our model the mechanism
works similarly for central and satellite galaxies.  Once galaxies are
embedded in a hot halo, they may continue to form stars until they
exhaust their existing reservoir of gas.  This allows some satellite
galaxies to survive as star-forming even as they fall through a hot
halo and merge with the central galaxy.  In some cases, as discussed
later, this leads to ``rejuvenation,'' where a satellite galaxy
delivers a new supply of star-forming gas to its central galaxy,
turning a quenched central into a star-forming one (for a brief time).
Rejuvenated galaxies generally grow very little in mass through the added
star-formation, but they can temporarily become green valley galaxies.

This simple hot gas quenching model produces a realistic $z=0$ red
sequence, with galaxy number densities matching local observations
\citep{gabor11}.  Central galaxies with stellar masses $\gtrsim
10^{10.5}$\Msun are the first to be quenched at $z>2$ because their
halos are dominated by hot gas at high redshift.  After joining the
red sequence they continue to accrete satellites and may grow in mass
by factors of 2 or more by $z=0$.  Satellites continually join these
massive halos and are also quenched, filling in lower masses on the
red sequence, especially at $z<1$.  Concurrently, new massive
star-forming galaxies are quenched as their halos grow to
$>10^{12}$\Msun down to $z=0$ \citep{gabor12}.

We note that in our simulations, mergers generally do not cause quenching.  Our
hot gas quenching model is not explicitly linked to galaxy mergers, and in our
simulations without hot gas quenching, mergers produce very few red
galaxies.  This result contrasts with some popular quenching models
\citep[e.g.][]{hopkins06_unified, hopkins08_ellipticals}.  In
\citet{gabor11} we showed that a separate quenching model based on
major mergers does not produce enough quenched galaxies, primarily
because merger remnants continue to accrete new fuel for
star-formation even if they eject all their pre-merger gas.

In \citet{gabor11} and \citet{gabor12} we identified some weaknesses of this
  quenching model.  Starvation due to the presence of hot gas alone
  has difficulty creating sufficient numbers of massive quenched
  galaxies at $z\gtrsim 2$.  This is partly because the timescale for
  gas starvation is a significant fraction of the age of the universe
  at very high $z$.  Thus a faster-acting quenching (such as galaxy
  mergers or violent disk instabilities) may be required in
  combination with starvation in the early universe.  Another problem
  is that our model with hot gas quenching produces too few massive
  star-forming galaxies, suggesting that this quenching is too
  efficient -- that is, hot gas quenching shuts down SF in \emph{all}
  massive galaxies, whereas a better model would shut down SF in
  \emph{most}, but not quite all such galaxies.  In simulations
  without quenching (but with stellar-driven winds), the high-$z$
  stellar mass function for all galaxies is in good agreement with
  observations.  The addition of a quenching model can only lower
  galaxy masses.  Thus, in order for a model with quenching to get
  enough high-mass, high-$z$ galaxies, the stellar feedback efficiency
  probably should be lowered so that massive galaxies can acquire more
  mass before they are quenched.  Despite these difficulties, hot gas
  quenching appears to be a good model for the low-$z$ red sequence.
  The problems noted above primarily apply to high-$z$ galaxies, while
  most of the red sequence is built up at $z<1$
  \citep[e.g.][]{faber07}. 

In summary, our quenching model prevents gas accretion onto galaxies
that live in regions dominated by hot gas, which typically arise
in halos with masses above $10^{12} M_\odot$.  Such galaxies are
effectively starved of new fuel for star-formation, and they fade to
red as they exhaust their remaining cold gas reservoir.

\subsection{Simulation runs}

For the remainder of the paper, we focus on simulations using our
quenching model (along with all the other physical processes above).
Our main simulation was run in a periodic cubic box with a side length
of $96 h^{-1}$~Mpc, with $512^3$ dark matter particles and $512^3$
initial gas particles.  This simulation allows us to model massive
structures such as several galaxy clusters with total $z=0$ masses a
few $\times 10^{14}$\Msun.  For convenience of analysis or clarity of
presentation, we also sometimes use smaller simulations with the same
resolution: we use both a $48 h^{-1}$~Mpc box with $2\times256^3$
particles, and a $24 h^{-1}$~Mpc box with $2\times 128^3$ particles,
as noted in the text.  We also make use of simulations with
  different sub-grid physics: we have already mentioned a simulation
  without quenching in \S\ref{sec.quench_model} (which was run in a
  $48 h^{-1}$~Mpc box), and we also use a simulation that includes
  neither quenching nor stellar-driven winds (also $48 h^{-1}$~Mpc).

The effective spatial resolution, set by the softening length, is
$3.75 h^{-1}$ kpc, and the mass of a gas particle is
$1.2\times10^8$\Msun.  We consider galaxies resolved if they contain
at least 64 star particles, or a stellar mass $\gtrsim 10^{9.5}$\Msun.
\citet{finlator06} showed that this particle number criterion
  leads to converged star-formation histories, at least for
  star-forming galaxies in similar simulations.  Our own convergence
tests with $\sim 3$ times better resolution than our main run
  (but which are run with smaller volumes and which do not proceed all
  the way to $z=0$) show that our results are not sensitive to
resolution. The hot halos that drive our quenching model are
sufficiently massive that they are always well-resolved.

All simulations use a WMAP5 cosmology with
$H_0=70$~km~s$^{-1}$~Mpc$^{-1}$, $\Omega_{\rm matter}=0.28$,
$\Omega_{\rm baryon}=0.046$, $\Omega_{\rm \Lambda}=0.7$ \citep{komatsu09}.  We
output $\approx 100$ simulation snapshots between $z=10$ and $z=0$,
with a time between consecutive snapshots of $\sim 100-300$~Myr.  For
most of this work, we focus on trends in the local universe, so we use
a snapshot at $z=0.025$ which is comparable to local observational
samples from e.g. SDSS \citep[][]{blanton05_vagc}.  We will typically refer to
this as $z=0$.

\subsection{Analysis methods}
\label{sec.analysis}
Once a simulation has run, we use a suite of tools to extract and
analyze galaxy properties.  We first identify galaxies and determine
basic properties (e.g. stellar mass) and derived properties
(e.g. absolute magnitudes).  Then we identify dark matter halos in
which the galaxies live, along with gas properties inside those
halos.  Finally, we quantify the large-scale environment around each
galaxy.

We identify galaxies using SKID, which relies on the DENMAX algorithm
to associate bound gas and star particles as galaxies.  The
galaxy stellar mass is the sum of the masses of the constituent
star particles, the star-formation rate is the sum of the
instantaneous SFRs of each gas particle, etc.  We determine galaxy
absolute magnitudes by considering each star particle as a single
stellar population with a given age and metallicity, assigning each
particle a spectrum based on a stellar population model \citep{bc03},
adding the spectra of all the star particles in the galaxy, and
convolving that spectrum with various filters (e.g. the SDSS ugriz
filters).  

We use ``red'' and ``quenched'' interchangeably to refer to galaxies
with low specific star formation rates in our simulations.  Simulated
galaxies are classified as either red or blue based on their locations
in an intrinsic absolute $g-r$ color-vs-stellar mass diagram -- we use
a mass-dependent color cut that varies with redshift
\citep[see][]{gabor11, gabor12}.  We do not model the reddening due to
dust, so galaxies on our red sequence are intrinsically red and thus
have little star-formation.  In fact, owing to our quenching model,
most of the red galaxies in our simulations have zero instantaneous
star formation rates.

We identify subhalos around each SKID galaxy, along with parent or
host halos, using a spherical overdensity algorithm.  We use
``subhalo'' to refer to matter that is within the virial radius of
each galaxy.  We use ``parent halo'' or ``host halo'' or just ``halo''
to refer to the largest halo in which a galaxy resides.  Every galaxy
has both a subhalo and a host halo -- for central galaxies, the two
are essentially the same.  We first associate every particle in a
simulation to the SKID galaxy to which it is most bound.  Then,
starting from each galaxy position, we iteratively step out in radius
and count up the mass of associated particles within the sphere
defined by that radius until the average density within that subhalo
falls below a redshift-dependent virial density appropriate for
our assumed cosmology as given in \citet[][$\approx 100\times$ the critical density]{dave10}.  This defines the virial
radius, and mass, of the subhalo.  We then ``merge'' subhalos.
Subhalos that lie within the virial radius of a larger subhalo are
subsumed into the larger halo -- the smaller subhalo is now considered
a satellite of the larger.  From the association of subhalos to larger
halos, we can identify groups and clusters of galaxies.

We measure the gas properties within each subhalo.  For the hot gas
fraction we simply add up the mass of hot gas within the subhalo
virial radius and divide by the total gas mass: $f_{\rm hot} = M_{{\rm
gas}, >10^{5.4} K} / M_{\rm gas}$.  This gives a measure of hot gas
fraction that is \emph{local} to each galaxy, and enables a
distinction between the local hot gas fraction around a satellite
galaxy and the hot gas fraction of the parent galaxy cluster in which it
lives.  We also measure radial profiles of hot gas fraction and gas
temperature for each subhalo.

We quantify galaxy environment using a 5th-nearest neighbor approach
akin to that used by observers \citep{dressler80, cooper05, cooper06,
kovac10, peng10}.  We use all resolved galaxies ($M_{\rm
stellar}>10^{9.5}$\Msun) to sample the density field.  The
overdensity at any point in the simulation box is $\delta = (\rho_5 -
\hat{\rho})/\hat{\rho}$, where $\rho_5 = 5 / ((4/3) \pi R_5^3)$, $R_5$
is the distance from the point to the 5th-nearest (resolved) neighbor
galaxy, and $\hat{\rho}$ is the average density of resolved galaxies
in the entire simulation box, $N_{\rm galaxies}/$(simulation volume).
In figures we generally plot $\log(1+\delta)$ and denote this quantity
as $\log({\rm overdensity})$.  Note that we use a 3-dimensional
5th-nearest neighbor, whereas observers are generally restricted to
projected measures.  Our tests show that trends remain unchanged by
this choice, but that a 3D measure better separates galaxies in the
densest environments \citep[see also ][]{cooper05}.

We trace the histories of halos and galaxies using a
progenitor-finding algorithm.  Consider two output snapshots, 1 and 2,
at redshifts $z_1$ and $z_2$ such that $z_1 < z_2$.  We seek the most
massive progenitor to halo $H_1$ at $z_2$.  For each dark matter
particle in halo $H_1$, we find the halo in which it lived at $z_2$
(if any).  Among these halos at $z_2$, the most massive one is
considered the main progenitor, $H_2$.  We follow a similar procedure
for star particles to identify the most massive progenitor of each
galaxy.  By comparing the $z=0$ snapshot to each previous snapshot, we
construct the history of each galaxy and halo from $z=0$ to $z\approx6$.


\section{Hot gas quenching trends}
\label{sec.trends}
%
%
\begin{figure*}
\includegraphics[width=168mm]{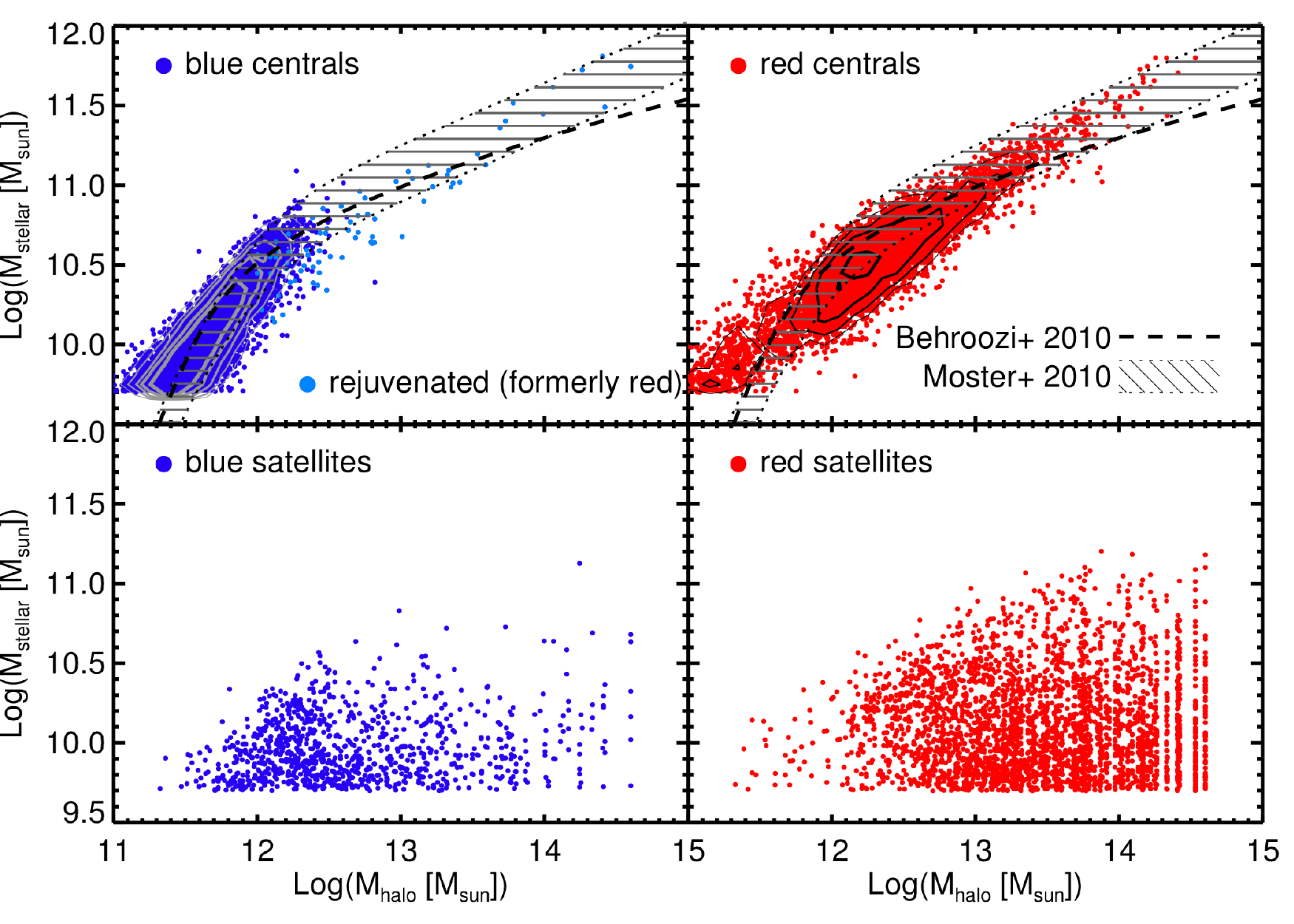}
\caption{Stellar mass as a function of host halo mass, separated into
  red/blue and centrals/satellites, as labelled in each panel.  For
  both blue centrals and red centrals (top row), stellar mass
  correlates strongly with halo mass, though the slope of the
  correlation for blue galaxies differs from that for red galaxies.
  The hatched region \citep{moster10} and dashed line \citep{behroozi10}
  show abundance matching constraints.  For
  satellite galaxies (bottom row), stellar mass does not correlate
  with halo mass.}
\label{fig.mstellar_mhalo}
\end{figure*}
%
\begin{figure}
\includegraphics[width=84mm]{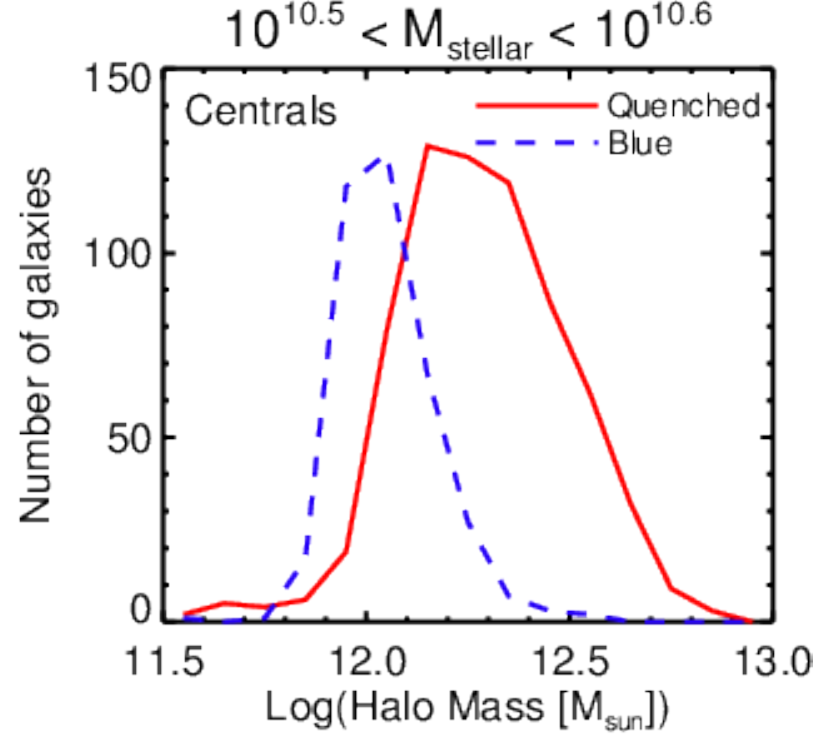}
\caption{Distribution of halo masses for blue star-forming (blue
  dashed line) and quenched red (red solid line) central galaxies in a
  narrow range of stellar mass, $10^{10.5} < M_{\rm stellar} <
  10^{10.6}$\Msun.  At fixed stellar mass,
  quenched galaxies live in more massive halos.}
\label{fig.mhalo_distribution}
\end{figure}

In this section we illustrate the crucial trends among stellar mass,
halo mass, environment, hot gas, and quenching that emerge from our
quenching model.  Many of these match the observed trends.  We
emphasize that both mass quenching and environment quenching emerge
naturally from our hot gas quenching model~\citep{gabor12}.

\subsection{The appearance of mass quenching and environment quenching}
%
Figure \ref{fig.2dhists} illustrates relationships among galaxy
stellar mass, environment, halo mass, hot gas fraction, and red
fraction in our model.  In the top left panel we show the fraction of
galaxies that are red (that is, quenched) as a function of environment
and stellar mass \citep[cf. Figure 7 of][]{gabor12}.  We compare
  with observational results by overplotting contours from an adjusted
  version of the fitting function from equation 7 of \citet{peng10}.
  Here overdensity is a 3-dimensional 5th-nearest-neighbor measure,
  whereas observations are restricted to projected measures.  We
  adjust the best-fitting critical overdensity that appears in the
  \citet{peng10} fitting function as follows.  First, we measure
  projected overdensities in our simulation \citep[as in][]{gabor12},
  then fit a line to the relationship between our projected and 3D
  measures.  Based on this best-fit line, the critical (logarithm of)
  overdensity of 1.84 becomes 2.57 when using a 3D density measure.
  The resulting contours are shown in Figure \ref{fig.2dhists}.

A characteristic boxy pattern emerges in the contours of red fraction,
identical to that arising in observations \citep{peng10}.  Galaxies
with high stellar masses are more likely to be quenched, independent
of environment, and galaxies in denser environments are more likely to
be quenched, independent of stellar mass.  Based largely on this boxy
pattern, \citet{peng10} argue that mass quenching and environment
quenching are separable, and therefore physically distinct processes.
Furthermore, some authors suggest that environment quenching is
decoupled from halo mass, since it appears to be independent of
stellar mass in this figure.

The top right panel of Figure \ref{fig.2dhists} shows the fraction of
galaxies whose subhalo is dominated by hot gas (with $f_{\rm hot} >
0.6$).  This panel shows the same boxy contours as that for red
fraction, exactly as expected since our quenching model is directly
tied to hot gas.

The bottom left panel of Figure \ref{fig.2dhists} shows the median
parent halo mass of galaxies in each bin of stellar mass and
environment.  $M_{\rm halo}$ generally increases with increasing
overdensity, but only at overdensities above $\sim 2$.  Below that
overdensity, $M_{\rm halo}$ generally increases with stellar mass.
Galaxies in massive groups and clusters with $M_{\rm
  halo}>10^{13}$\Msun typically live in the highest-density regions,
but some massive galaxies live in $\sim10^{13}$\Msun groups at low
overdensities.  That is, galaxies in halos with $M_{\rm halo}\approx
10^{13}$\Msun live in the full range of environments.  We return to
this point in \S\ref{sec.MQ_env}.

The bottom right panel of Figure \ref{fig.2dhists} shows the parent
halo mass in a different way -- it shows the fraction of galaxies
whose parent halo is above $10^{12}$\Msun.  Here, a similar boxy
pattern emerges as in the top two panels.  Thus, the appearance of
separable mass quenching and satellite quenching can emerge as a
direct consequence of a critical halo mass (here $\sim 10^{12}$\Msun) where
galaxies become quenched.  

In summary, quenching in hot, massive halos naturally leads to the
appearance of both mass quenching and environment quenching.  The two
quenching modes are in fact manifestations of the same process in our
model.  Despite our unified quenching mechanism, these appear to be
mathematically separable as defined by \citet{peng10}.  The
separability arises because massive halos which host massive central
galaxies ( e.g. $M_{\rm halo} > 10^{12}$ for $M_{\rm stellar}>
10^{10.5}$\Msun) inhabit nearly the full range of environments, and
galaxies in the densest environments may have a wide range of stellar masses.

\subsection{Mass quenching is central quenching; Environment quenching is satellite quenching}
The observed mass quenching is mainly thought to be a result of
quenched central galaxies, whereas environment quenching is the result
of quenched satellites (e.g. \citealt{peng12, woo13}; see also
e.g. \citealt{delucia12b}).  Figure \ref{fig.fsat} shows
that this is the case for our model.  The galaxies with high red
fraction at low $M_{\rm stellar}$ and high overdensity (i.e. towards
the upper left of the figure) are satellite galaxies.  Red galaxies
with high $M_{\rm stellar}$ but low overdensities (towards the lower
right) are centrals.  Massive galaxies in dense environments (upper
right) can be either centrals or satellites.

\subsection{$M_{\rm stellar}-M_{\rm halo}$ relations}
The stellar mass-halo mass relation is widely used to quantify the
efficiency with which halos of different masses convert their baryons
into stars \citep[e.g.][]{mandelbaum06, conroy07_halo_mass}.  It is
closely linked to stellar mass function measurements through
abundance-matching techniques \citep[e.g.][]{moster10, behroozi10},
and has been well-studied especially in the $z<1$ universe with
additional constraints from galaxy lensing and clustering
\citep[e.g.][]{leauthaud12a, leauthaud12b, tinker13}.  Under the
  assumption that our understanding of dark matter structure formation
  is good, the stellar mass-halo mass relation serves as an important
  constraint on baryonic physics.

In our hot gas quenching model, star-forming central galaxies build up
both halo mass and stellar mass over time as they accrete dark matter
and gas.  During this phase, galaxies accrete intergalactic gas
without the hindrance of a hot halo, although their efficiency of converting
baryons into stars is suppressed in a mass-dependent manner owing to
star formation-driven outflows \citep{dave12}.  Once a galaxy's halo mass
is above about $10^{12}$~\Msun, its halo's hot gas fraction is likely
to be above 60\% (Figure \ref{fig.hotfrac_vs_mass}), and its fuel
supply for star-formation is cut off in our quenching model.  In the
simulations, as in observations \citep[e.g][]{peng12}, halo mass is
tightly correlated with galaxy stellar mass for central galaxies.  We
show this relation in the top panels of Figure
\ref{fig.mstellar_mhalo}.  The $M_{\rm stellar}-M_{\rm
halo}$ relation is tight for both blue and red centrals, but the
slope differs for these two populations. For comparison, we also show observationally-inferred relations from \citet{moster10} and \citet{behroozi10}.

Based on this correlation, a central galaxy with a halo mass of
$10^{12}$~\Msun has a stellar mass of $\approx10^{10.5}$~\Msun.
Galaxies above this stellar mass tend to be quenched.  Some central
galaxies below this mass are also red; we address these in
\S\ref{sec.lowmass_red}.  Quenched galaxies above this stellar mass
continue to grow via dry or semi-dry mergers \citep{gabor12}, and they
do so in such a way that maintains the correlation with halo mass --
centrals in more massive halos accrete more stellar mass via mergers.
We also note that these massive galaxies may accrete gas-rich
satellites.  This may lead to a brief ``rejuvenation'' of
star-formation in the central galaxy such that it once again appears
blue.  Such cases are marked in the ``blue centrals'' (top left) panel
of Figure \ref{fig.mstellar_mhalo} -- these galaxies were previously
quenched (for at least 3 consecutive snapshots) before becoming blue
again.  Though rare, these rejuvenated blue galaxies appear to follow
the slope of \emph{red} centrals in the $M_{\rm stellar}-M_{\rm
halo}$ relation, indicating that the rejuvenation does not lead to
much stellar mass growth.  The effective critcal stellar mass $\sim
10^{10.5}$~\Msun, where the $M_{\rm stellar}-M_{\rm halo}$ relation
changes slope, is closely related to the ``knee'' the in galaxy
stellar mass function, as shown in \citet{gabor12}.

The presence of hot gas in massive halos therefore leads to the
appearance of ``mass quenching'' among central galaxies: as galaxies
grow more stellar mass, they also grow more halo mass; once they reach
sufficiently high halo mass, they become hot gas-dominated; and in our
quenching model, the hot gas quenches their star-formation.  The
efficiency of stellar mass growth (relative to the halo mass)
decreases for these high mass galaxies because star-formation is
suppressed.

In the bottom panels of Figure \ref{fig.mstellar_mhalo}, we show the
$M_{\rm stellar}-M_{\rm halo}$ relation for blue and red satellites.
For satellite galaxies, the stellar masses are essentially
uncorrelated with halo masses.  There is an upper envelope of stellar
masses (there are no satellites towards the upper left of these
panels) which arises simply because satellites cannot have masses
larger than the centrals of their host halos \citep[as pointed out in
  observational samples by][]{peng12}.  Thus the stellar masses of
satellite galaxies are mostly unrelated to their $z=0$ parent halo
masses.

In summary, stellar masses are closely correlated with halo masses for
central galaxies, but not for satellites.  The slope of the $M_{\rm
  stellar}-M_{\rm halo}$ relation for centrals changes at the
characteristic quenching mass scale, $M_{\rm halo}\approx 10^{12}$\Msun or
$M_{\rm stellar}\approx 10^{10.5}$\Msun.

\begin{figure*}
\includegraphics[width=168mm]{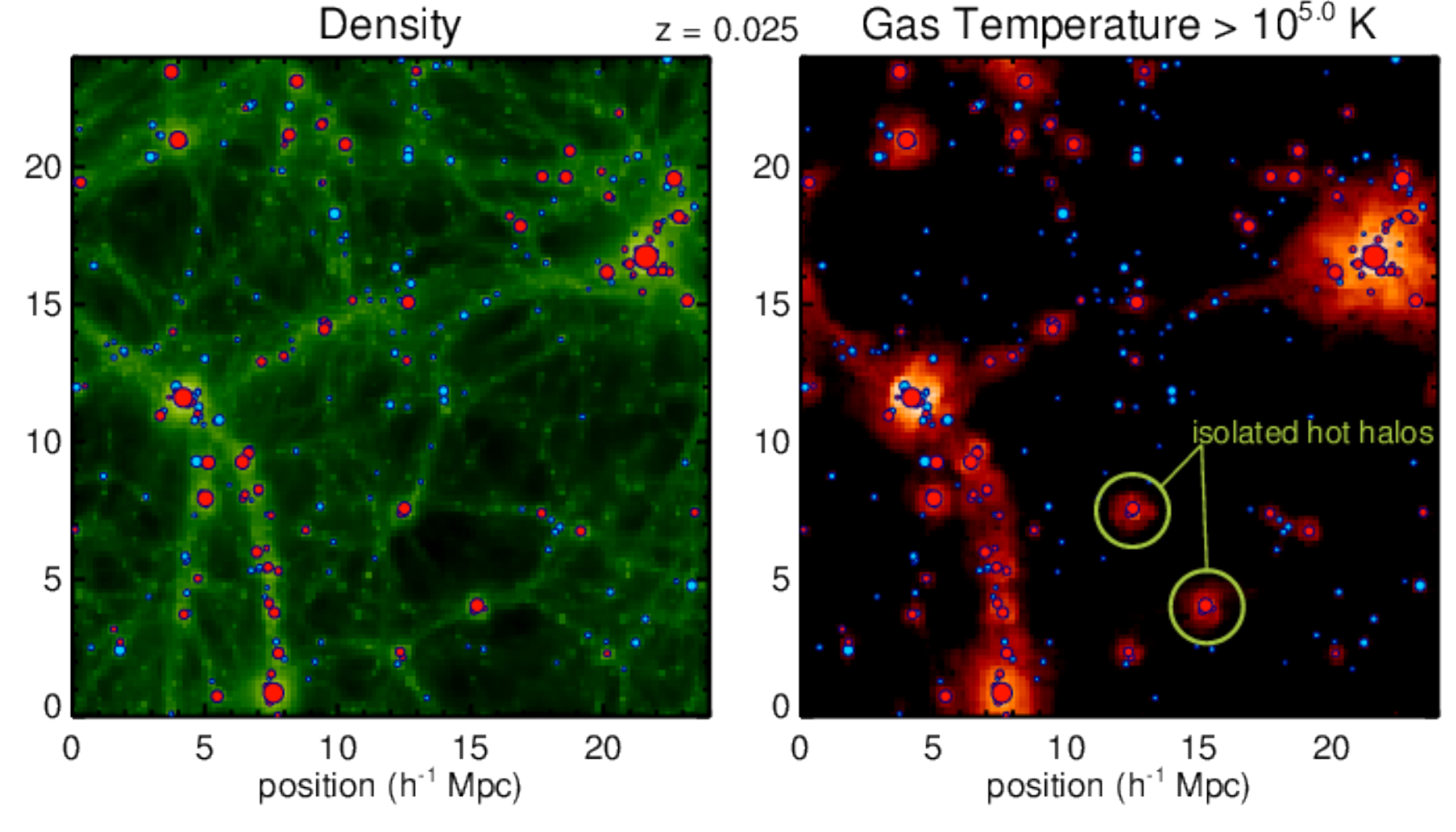}
\caption{Density (left) and temperature (right) images of gas in the
  cosmic web, with galaxies overplotted as circles.  The color of each
  point indicates whether the galaxy is red or blue, and the size of
  the circle scales with the square root of stellar mass.  In the
  right panel, gas is only shown (in red scale) if it is hot
  ($>10^{5.4}$~K).  Red galaxies tend to live in more clustered
  environments, but red centrals in hot halos are sometimes isolated.
  We highlight two examples of relatively isolated hot halos with
  large green circles.}
\label{fig.gas_im}
\end{figure*}
\begin{figure}
\includegraphics[width=84mm]{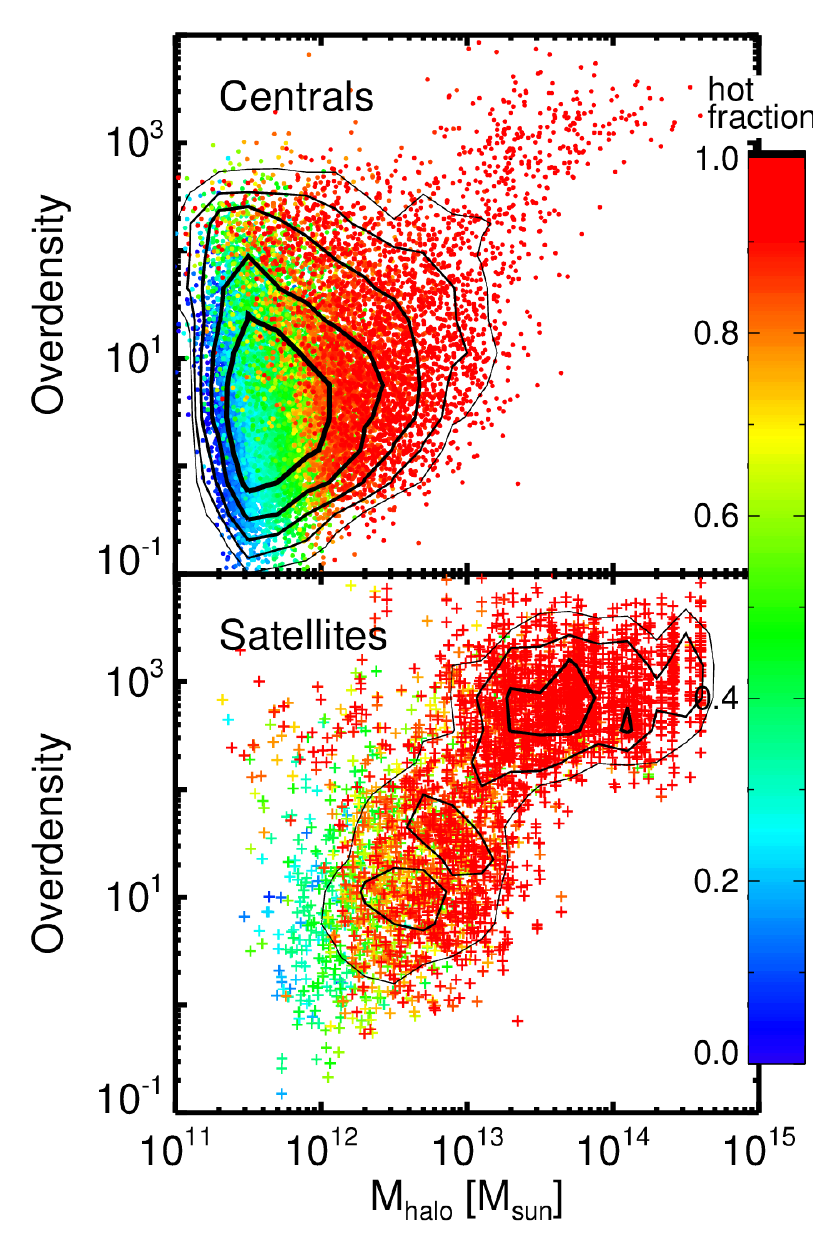}
\caption{Environment vs. parent halo mass for central galaxies (top)
  and satellite galaxies (bottom).  The color of each symbol indicates
  the hot gas fraction in the local vicinity of each galaxy.  For
  centrals, environment is poorly correlated with halo mass, except in
  the rare, most massive halos.  Thus halos with $10^{12}<M_{\rm
    halo}<10^{13}$\Msun, which are generally dominated by hot gas, span
  nearly the full range of environments.  Likewise, overdensity serves
  as a poor indicator of halo mass for centrals. For satellite
  galaxies the correlation between overdensity and halo mass is more
  clear, although with significant scatter.}
\label{fig.env_mhalo}
\end{figure}
%
%

\subsubsection{At fixed $M_{\rm stellar}$, red central galaxies live in more massive halos than blue galaxies}
\label{sec.halo_bias}
In Figure \ref{fig.mhalo_distribution} we show histograms of halo mass
for blue and red central galaxies in a narrow range of stellar mass.
This is like taking a narrow horizontal slice through the top panels
of Figure \ref{fig.mstellar_mhalo}, and plotting the halo mass
distributions.  We chose a mass range $10^{10.5} < M_{\rm stellar} <
10^{10.6}$\Msun because there are roughly equal numbers of blue and
red centrals.  This stellar mass is also near the ``knee'' in the
simulated stellar mass function \citep{gabor12}.  The red galaxies
live in significantly more massive halos than their star-forming
counterparts.  This qualitatively agrees with clustering-based
analyses, at least in the local universe \citep[e.g.][]{tinker13}, and
dynamical halo mass estimates \citep{phillips14}.  This ``bias'' in
halo masses at fixed stellar mass may help explain galactic conformity
-- the tendency for quenched centrals to have quenched satellites, and
star-forming centrals to have star-forming satellites
\citep{weinmann06, kauffmann13, robotham13, phillips14}.  In our model, the
quenched centrals live in more massive, hot halos, so their satellites
are more likely to be quenched.  We leave a more detailed analysis of
galactic conformity for future work.

We can explain this effect physically two ways.  (1) Immediately after
a star-forming galaxy is quenched, its dark matter halo may continue
to grow in mass while its stellar mass remains the same (ignoring
galaxy mergers).  (2) For main sequence star-forming galaxies, there
is scatter in the $M_{\rm stellar}-M_{\rm halo}$ relation.  Those
galaxies in the high-halo mass tail of this scatter are more likely to
be quenched in our model.  Thus the quenched galaxies will have higher
halo mass.  In practice both of these effects may combine to produce
the difference in halo mass at fixed stellar mass.

Finally, we note (without showing in Figure
\ref{fig.mhalo_distribution}) that this trend is reversed for low-mass
central galaxies: for $M_{\rm stellar}<10^{10}$\Msun, red galaxies
actually live in less massive dark matter halos than their blue
counterparts (though there are many more blue galaxies at these
masses).  As we explain in \S\ref{sec.lowmass_red}, such low-mass
quenched central galaxies are predominantly ejected former satellites
of much more massive halos.  Thus their low halo masses could be
explained by stripping of the dark matter halo through interaction
with their former host halos.


\subsection{Mass quenching independent of environment}
\label{sec.MQ_env}
Based on the boxy contour shape in Figure \ref{fig.2dhists}, mass
quenching appears to be mostly independent of environment.  In this
section, we emphasize a simple explanation for this fact in light of
our hot gas quenching model: hot halos (broadly, those with $M_{\rm
halo}\ga 10^{12}$\Msun) occupy the full range of environments.

Although the critical halo mass for forming hot gas ($10^{12}$~\Msun)
is fairly large, such hot halos live in a wide variety of environments
from the field to large groups.  We show this visually in Figure
\ref{fig.gas_im} (using our $24h^{-1}$~Mpc simulation for
  clarity).  The left panel shows the cosmic web density, and the
  right panel shows gas temperature only for hot gas ($>10^{5.4}$~K).
  Over each image we plot the positions of red and blue galaxies (the
  size of each galaxy point is proportional to the square root of its
  stellar mass).  As expected, the densest regions host lots of hot
  gas and red galaxies, while less dense filaments generally host
  star-forming galaxies.  But some hot halos form in relatively
  isolated regions, as indicated in the figure.  These isolated hot
  halos (which typically have masses just above $10^{12}$~\Msun)
  generally host red galaxies.

For a more quantitative look, we show the relationship between parent
halo mass and environment in Figure \ref{fig.env_mhalo}.  For central
galaxies, the correlation is poor except in the most extreme, massive
halos.  Around $10^{12}$~\Msun, central
galaxy halos span nearly the full range of environments, so
environment is not a good indicator of halo mass.  This explains why
mass quenching appears effectively independent of environment in both
our simulations and in observations \citep{peng10}.

For satellite galaxies (lower panel of Figure \ref{fig.env_mhalo}) the
relationship between parent halo mass and environment is more clear.
Satellites living in massive groups are likely to have many nearby
neighbors.  Environment is therefore a reasonbly good indicator of parent halo
mass for satellites, although large scatter remains.

In summary, in our hot gas quenching model, halos around the critical
quenching halo mass of $10^{12}$\Msun live in the full range of
environments.  Thus environment is not a good indicator of parent halo
mass for quenching halos, and mass quenching/central quenching appears
to be independent of environment.  Satellite galaxies, on the other
hand, show a correlation between parent halo mass and environment.

\subsection{Environment/Satellite quenching appears to be independent of $z=0$ halo mass}
%
%
%
\begin{figure}
\includegraphics[width=84mm]{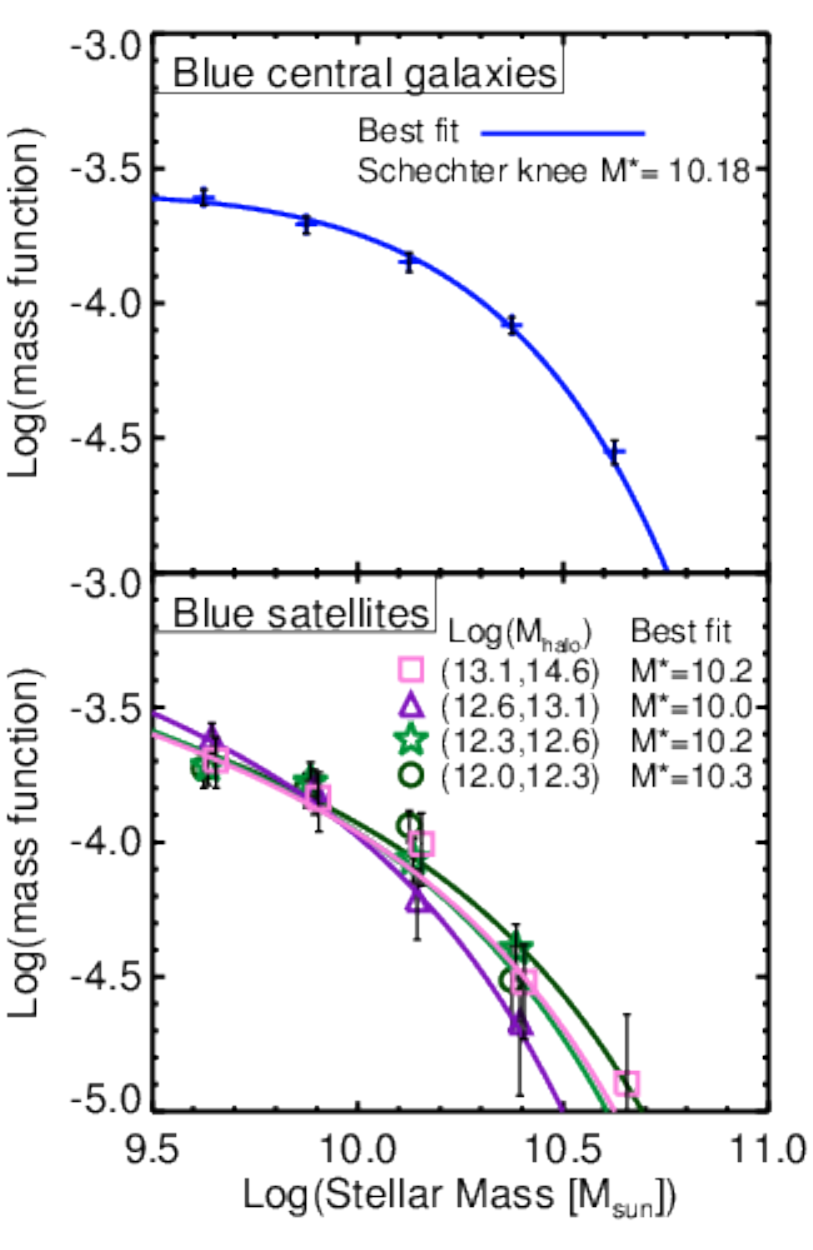}
\caption{Stellar mass functions (number density of galaxies per
  logarithmic stellar mass bin) of blue centrals ({\bf top}) and blue
  satellite galaxies in 4 bins of halo mass (chosen so that each bin
  has roughly the same number of galaxies, {\bf bottom}).  Symbols
  show the actual mass functions, while lines show Schechter function
  fits with the faint-end slope fixed to -1.4 and the knee $M^*$ free
  to vary (the normalization is arbitrary; \citealt{peng12}).  The
  best-fit critical mass $M^*$ shows no trend with halo mass, implying
  that mass quenching of satellite galaxies occurs independent of halo
  mass, as in observations.}
\label{fig.bluesat_mf}
\end{figure}

While central galaxies are quenched when their halos become massive
enough to support a hot atmosphere, two paths are possible for
quenched satellite galaxies -- 1) they are quenched as centrals before they
become satellites, then merge into a larger halo, or 2) they are
quenched as satellites by the hot gas existing in their parent halos.
The second leads to the appearance of ``environment quenching,'' in
which most galaxies in dense regions are red (cf. Figure
\ref{fig.2dhists}).  This satellite quenching mechanism is mostly
independent of stellar mass because galaxies of all stellar masses may
be satellites in quenching halos above $10^{12}$~\Msun.

\citet{peng12} noted that in SDSS data, environment or satellite
quenching appears to be independent of the parent halo mass of
satellite galaxies.  In particular the knee, $M^*$, in the stellar
mass function of blue satellite galaxies is independent of halo mass,
and is the same as that for blue central galaxies.  This indicates
that blue satellites are subject to mass quenching at the same
critical stellar mass, regardless of halo.  These observed trends
emerge from our quenching model, despite the implicit
dependendence of quenching on halo mass (cf. Figure
\ref{fig.hotfrac_vs_mass}).



In the top panel of Figure \ref{fig.bluesat_mf} we show the stellar
mass function of blue central galaxies.  Symbols with error bars show
the measured stellar mass function from our simulation (with arbitrary
normalization).  A single-Schechter fit to the blue central galaxies
(shown as a solid line) yields a knee in the mass function of
$\log(M^*/$\Msun$)=10.18$.  This value for $M^*$ is notably smaller
than that in observations \citep[cf.][]{peng10, ilbert13, moustakas13}
-- our simulations likely produce too few massive blue galaxies
\citep{gabor12}, as noted in \S\ref{sec.quench_model}.  Mass quenching
sets this $M^*$ cutoff because galaxies more massive than $M^*$ are
increasingly likely to live in massive ($>10^{12}$\Msun), hot halos
and thus be quenched.

In the bottom panel of Figure \ref{fig.bluesat_mf} we show stellar
mass functions in 4 bins of parent halo mass.  We choose the 4 halo
bins to contain roughly the same number galaxies.  When splitting the
sample, the errorbars become substantial since the number of
galaxies per bin is fairly small.  Following Peng et al., we fit each
stellar mass function with a single Schechter function with a power
law slope fixed to -1.4.  In these fits the normalization is
arbitrary, and we allow $M^*$ to vary.  We find that the knee
$M^*$ shows no trend with halo mass, with all halo mass bins within
0.2 dex of the median $log(M^*/$\Msun$)=10.2$.

In summary, mass quenching applies to satellite galaxies independent
of parent halo mass.  The stellar mass function ``knee'' (i.e. the
critical quenching mass) for blue satellites is independent of halo
mass, and it is the same as that for blue centrals.  This independence
on halo mass appears despite the close link with halo mass in our hot
gas quenching model (cf. Figure \ref{fig.hotfrac_vs_mass}).  In the
next section, we present a physical explanation for this -- many satellite
galaxies are pre-processed before entering their $z=0$ halos.

\subsection{Pre-processing of satellite galaxies}
\begin{figure}
\includegraphics[width=84mm]{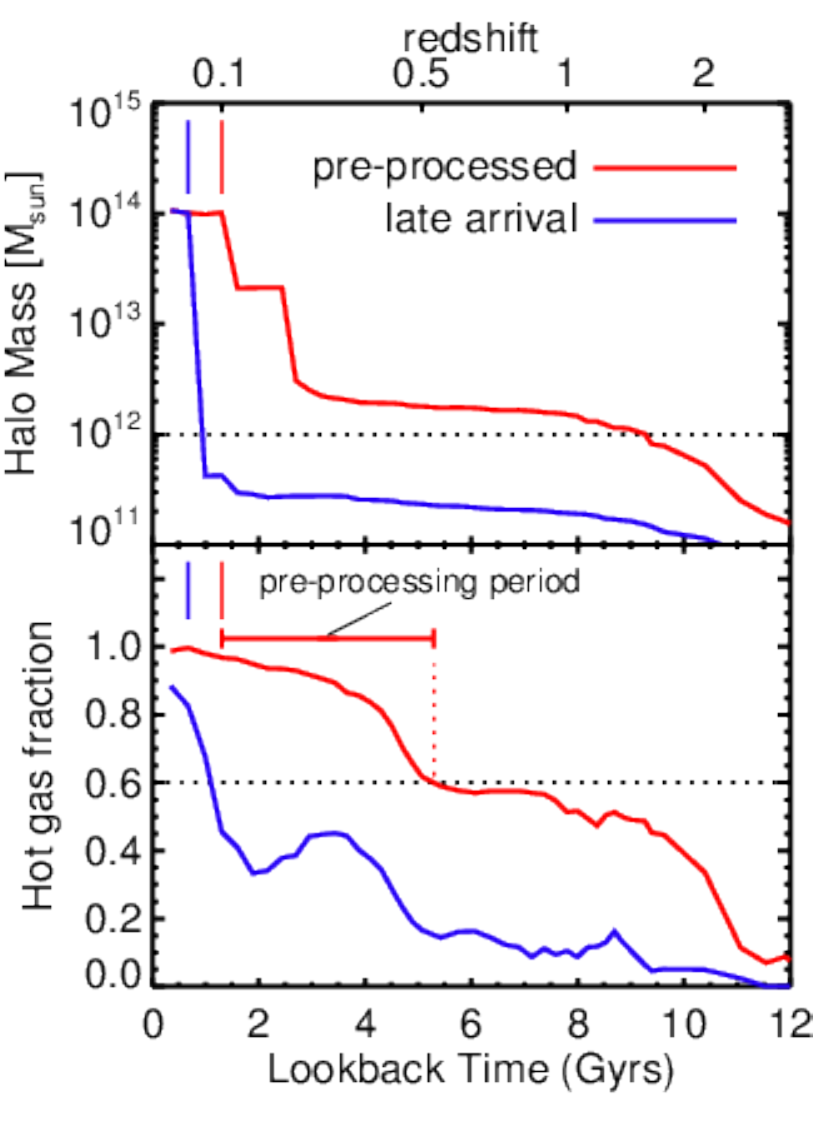}
\caption{{\bf Top:} Parent halo mass as a function of time for two
  galaxies that end up as satellites in the same galaxy cluster at
  $z=0$.  Short vertical lines (top left of each panel) indicate when
  each galaxy merged with the cluster.  One galaxy is pre-processed in
  other halos with masses $>10^{12} $\Msun (red line), while the other
  lives only in low-mass halos (blue line) before joining the
  cluster. {\bf Bottom:} Hot gas fraction (measured with the local subhalo)
  as a function of time for the same two galaxies.  The pre-processed
  galaxy lives in a hot environment (above the critical hot fraction
  of 0.6, dotted line) starting at $z\approx 0.5$ -- it is
  pre-processed in this hot gas for about 4~Gyr before finally merging
  with the cluster at $z\approx0.1$.  The ``late arrival'' galaxy does
  not live in hot gas until it merges with the cluster.}
\label{fig.halo_hist}
\end{figure}
\begin{figure}
\includegraphics[width=84mm]{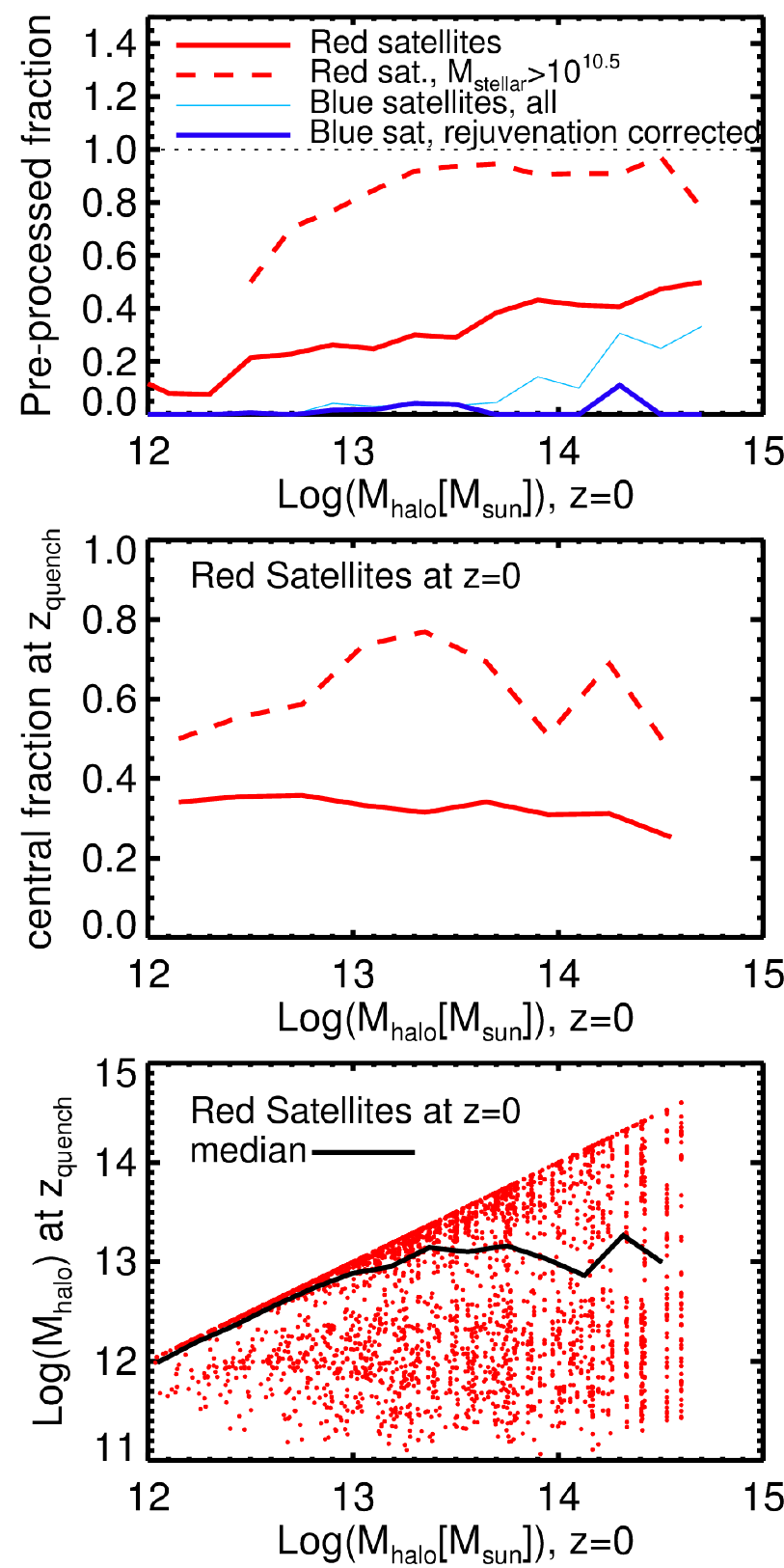}
\caption{{\bf Top:} Fraction of galaxies pre-processed in group-sized
  halos as a function of $z=0$ halo mass.  In cluster-mass halos,
  roughly half of red satellites (solid red) and nearly all massive
  red satellites (dashed) were processed in groups before joining
  their $z=0$ halo.  Most blue satellites (solid blue lines) that were
  pre-processed have actually been rejuvenated via gas-rich minor
  mergers -- after excluding rejuvenated galaxies (thick, dark
  blue line), almost no blue satellites have been pre-processed. {\bf
    Middle:} Fraction of $z=0$ satellite galaxies that {\it were}
  centrals at the redshift when they were quenched, $z_{\rm quench}$,
  as a function of $z=0$ host halo mass.  {\bf Bottom:} Host halo mass
  of red satellites when they were quenched as a function of $z=0$
  host halo mass.  In the median, satellites in massive groups and
  clusters at $z=0$ lived in a halo of $\sim 10^{13} $\Msun when they
  were first quenched (with a large scatter).}
\label{fig.fprocess}
\end{figure}
How can a quenching process closely linked to halo mass lead to an
apparent lack of dependence on halo mass? The answer is, through 
pre-processing -- satellite galaxies in clusters at $z=0$ may have been 
pre-processed in groups at higher redshifts.  Galaxies may assemble into 
groups at higher redshifts, and if their group masses are above $\sim10^{12}
$\Msun, then they will be dominated by hot gas and begin quenching.
Later, after the group galaxies have been quenched, the group may
merge with a galaxy cluster.  The group central galaxy, which has
effectively been ``mass quenched,'' becomes a satellite galaxy of the
cluster.

Several previous authors have studied pre-processing in simulations
\citep{berrier09, mcgee09, delucia12b}.  They typically identify
galaxies in cluster-mass halos ($>10^{14}$\Msun) that have been
pre-processed in what we call massive groups, with $M_{\rm
  halo}>10^{13}$\Msun.  A more natural mass scale for pre-processing
halos is that of ``small groups,'' $M_{\rm halo}=10^{12}$\Msun \citep{williams12}.  This
is the mass scale where the virial shock forms a stable hot halo
(cf. Figure \ref{fig.hotfrac_vs_mass}), and in our simulations these are the
halos where quenching begins.  In what follows, we use $10^{12}$\Msun
as the pre-processing halo mass -- galaxies in groups or clusters at
$z=0$ may have been pre-processed in halos above $10^{12}$\Msun at
earlier epochs.  This obviously includes the central galaxies of $\sim
10^{12}$\Msun halos, which some authors consider field galaxies. Below
we further define additional criteria to determine whether a galaxy
has been pre-processed.

Figure \ref{fig.halo_hist} illustrates an example of pre-processing in
our simulations.  We show halo mass histories (top) and hot gas
fraction histories (bottom) for two galaxies that end up in the same
cluster at $z=0$.  One galaxy is pre-processed, while the other
is not -- it is a ``late arrival'' to the cluster environment.
The pre-processed galaxy has lived in halos $>10^{12} $\Msun since
before $z=1$, and it lived in a hot gas-dominated halo since $z\approx
0.5$.  Thus it was quenched well before joining the cluster.  On the
other hand, the late arrival galaxy lived only in low-mass, cold halos
until joining the (hot) cluster at about 1~Gyr lookback time.  At
$z=0$, although it accretes no new gas, this galaxy continues to form
stars at a low level because it has not yet exhausted its supply of
cold gas.

Based on halo mass and hot gas histories like these, we determine
whether each satellite galaxy at $z=0$ was pre-processed.  This
requires tracing the halo mass history of each galaxy across cosmic
time, as well as the full history of the halo in which the galaxy ends
up at $z=0$.  We will refer to the latter -- the $z=0$ host halo and
its most massive progenitors -- as the ``main line'' halo (this may be thought of
as the ``trunk'' in a merger tree).  Galaxies generally start out in
distinct halos then merge with the ``main line'' halo to add to the
cluster or group.  A satellite is considered pre-processed if it
satisfies the following three criteria: (1) at some time in the past,
the galaxy lived in a halo distinct from the main line halo; (2) the
parent halo just before merging with the main line halo had a lower
mass than the main line halo, and a higher mass than $10^{12}$\Msun;
(3) the galaxy lived in a hot gas-dominated environment for at least 1
Gyr prior to merging with the main line.  These criteria ensure that
the galaxy lived in a separate hot group before joining the final,
larger group or cluster.

In the top panel of Figure \ref{fig.fprocess} we show the fraction of
satellites that were pre-processed as a function of $z=0$ parent halo
mass.  The pre-processed fraction of red satellite galaxies (solid red
line) is a slowly increasing function of halo mass, ranging from
$20-50$ per cent.  Roughly half of cluster satellites (i.e. in halos
$>10^{14} $\Msun) have been pre-processed before joining the cluster.
This includes massive red satellites (approximately, above knee in the
stellar mass function, $M_{\rm stellar} > 10^{10.5}$\Msun), of which a
vast majority were pre-processed.

The pre-processed fraction for blue satellites (blue lines) is very
low.  In cluster halos up to 30 per cent of blue satellites may be
pre-processed, but most of these galaxies turn out to be rejuvenated
galaxies -- they were previously quenched, but they are forming stars
because minor mergers with gas-rich satellites have provided a small
amount of new fuel for star formation.  Excluding rejuvenation (thick,
dark blue line), a large majority of blue satellites have moved
directly from ``the field'' to their final group or cluster
environment.  These blue galaxies are typically on their way to being fully
quenched, but have not yet exhausted their fuel.

In the middle panel of Figure \ref{fig.fprocess} we show the fraction
of galaxies that were centrals at the time they were first quenched.
We define $z_{\rm quench}$ as the highest redshift where a galaxy was
red for at least two consecutive output snapshots.  Roughly 30 per
cent of $z=0$ red satellites were centrals when first quenched, including
$50-75$ per cent of massive red satellites.  This means that most
massive satellites were originally ``mass quenched'' in their own
halos rather than ``environment quenched'' as satellites of a cluster.  

In the bottom panel of Figure \ref{fig.fprocess} we show, for red
satellites, the parent halo mass at $z_{\rm quench}$ as a function of
parent halo mass at $z=0$.  Especially for massive groups and clusters
($M_{\rm halo}>10^{13}$\Msun) at $z=0$, a large fraction of red satellite
galaxies lived in substantially lower-mass halos when they were first
quenched.

In summary, $\sim 20-50$ per cent of red satellite galaxies were
pre-processed in lower-mass hot halos before joining their $z=0$ halo;
most blue satellites were not pre-processed.  Most massive red
satellites (with $M_{\rm stellar} > 10^{10.5}$\Msun) at $z=0$ were
pre-processed as centrals galaxies at earlier times.  This helps
explain why satellite quenching does not show strong trends with halo
mass -- the knee in the mass function is largely set by ``mass
quenching'' which occurs before the satellites arrive in their final
halos.

\subsection{Quenching and distance to the center of the parent halo}
\label{sec.dis}
\begin{figure}
\includegraphics[width=84mm]{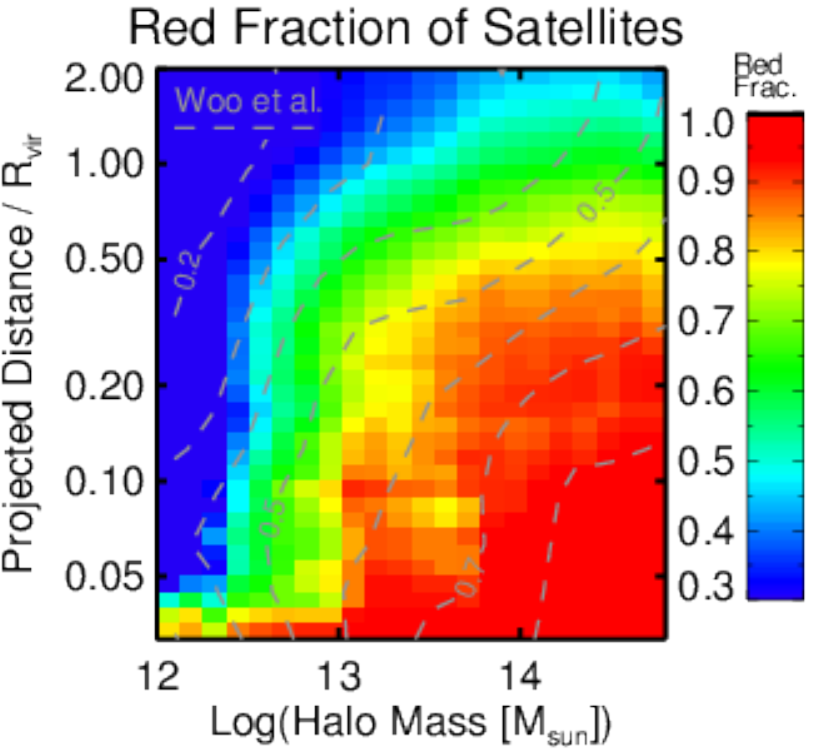}
\caption{Red (i.e. quenched) fraction as a function of projected
  distance from the halo center and parent halo mass, for satellite
  galaxies.  Here, we extend out to twice the virial radius, and count
  any galaxies within this doubled radius (in projection) as
  satellites.  The shape of the contours is qualitatively similar to
  observations from \citet[][grey dashed contours]{woo13}. In massive
  groups and clusters, $\sim 50$ per cent of satellites are quenched
  at the virial radius.  This suggests that the hot gas responsible
  for quenching extends beyond the virial radius.}
\label{fig.woo_dist_mhalo}
\end{figure}
%

\begin{figure}
\includegraphics[width=84mm]{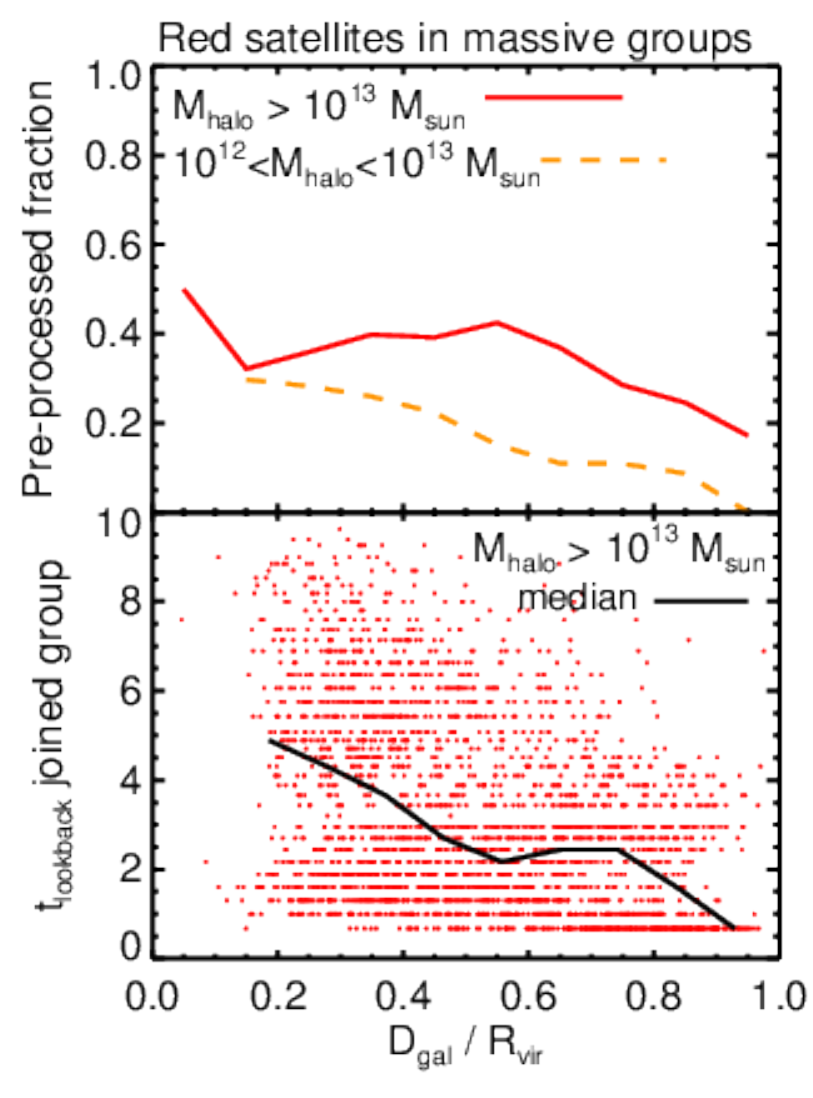}
\caption{{\bf Top:} Pre-processed fraction vs. distance (relative to
  $R_{\rm vir}$ for satellites in massive groups and clusters (halos
  above $10^{13}$\Msun, solid line).  The pre-processed fraction
  varies weakly with distance -- galaxies near the centers of halos
  are just as likely as those at large radii to be
  pre-processed. Low-mass groups (dashed line) have fewer
  pre-processed satellites, and show a stronger radial gradient.  {\bf
    Bottom:} Lookback time (in Gyr) at which satellites joined their
  $z=0$ group or cluster (halos $>10^{13}$\Msun), as a function
  of distance.  Satellites near the edges of halos have typically
  joined more recently, although there is large scatter.  Some
  galaxies near the centers of halos have merged with their host
  groups or clusters within the last 2 Gyr.}
\label{fig.dis_fprocess}
\end{figure}
Based on observations, \citet{woo13} argue that distance from the halo
center is a better predictor of quenching than environment for
satellite galaxies.  In Figure \ref{fig.woo_dist_mhalo} we show the
red fraction as a function of apparent projected distance from the
halo center (normalized to the virial radius, i.e. $D_{\rm gal} /
R_{\rm vir}$) and halo mass.  For massive groups and clusters in our
simulations (halos with $M_{\rm halo}>10^{13}$\Msun), it turns out
that a significant fraction of satellites are quenched even beyond $1~
R_{\rm vir}$.  Thus, we include as ``satellites'' in this figure any
galaxy that appears within two projected $R_{\rm vir}$.  If a galaxy
is within the virial radius of two different halos, it is assigned to
the more massive halo.

The resulting contours in Figure \ref{fig.woo_dist_mhalo} show that
the quenched fraction declines both as halo mass decreases and
distance increases.  These contours broadly match the shape of those
in Figure 10 of \citet{woo13}, which we show here as grey dashed
  lines.  In detail, the contours from our simulation lack some of the
  structure shown in the observations, and the simulated quenched
  fraction is higher overall.  This suggests our quenching model is
  somewhat too efficient in these massive halos (as discussed in
  \S\ref{sec.quench_model}).  In simulated massive groups and
clusters, greater than half of satellite galaxies are quenched at the
virial radius (which is probably too high; \citealt{wheeler14}), and
many nearby galaxies are quenched even if they are beyond the virial
radius.  Nevertheless, the overall trend of quenching with halo mass and distance is
striking, and qualitatively similar to that in observations
\citep{wetzel12}.

In summary, satellite galaxies are more likely to be quenched when
they are closer to their parent halo's center, and when they live in
more massive halos.  In our models many galaxies are quenched beyond
the virial radius of massive halos, and we further explore this effect
below in \S\ref{sec.hot_beyond}.

\subsubsection{Radial positions of pre-processed galaxies} 
Now we briefly return to pre-processing and the joining of satellites into
massive clusters and groups.  In the top panel Figure
\ref{fig.dis_fprocess} we show the pre-processed fraction as a function of
distance from the halo center, for red satellites in massive groups
and clusters with $M_{\rm halo} >10^{13}$\Msun.  The pre-processed
fraction varies weakly with radius -- even in cluster cores, $>30$ per
cent of satellites are pre-processed.  Some of these central
pre-processed galaxies may have plunged toward the cluster center at
late times, but many were pre-processed at high redshift before the
bulk of the cluster assembled.

In the bottom panel of Figure \ref{fig.dis_fprocess}, we show the lookback
time at which each satellite merged into its final, $z=0$ halo, as a
function of the distance from the halo center.  As naively expected,
galaxies at the outskirts of a group or cluster have merged with it
more recently than those in the core.  There is, however, a great deal
of scatter, with many galaxies in cluster cores having joined the cluster
within the last 2 Gyrs.  Massive groups and clusters are dynamically
complex, with some satellites able to plunge to the cluster core soon
after merging with the halo.

\subsection{Hot gas beyond the virial radius}
\label{sec.hot_beyond}
\begin{figure}
\includegraphics[width=84mm]{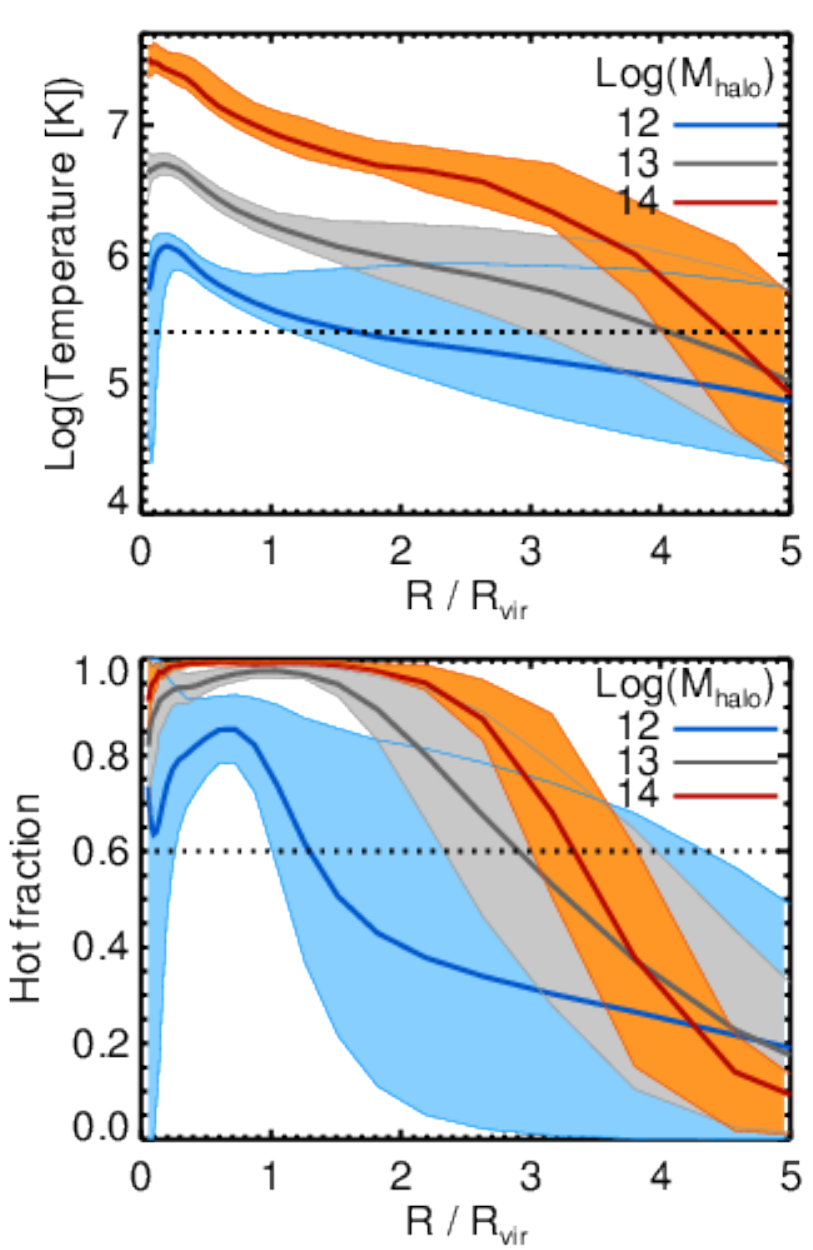}
\caption{Gas
  temperature ({\bf top}) and hot gas fraction ({\bf bottom})
  vs. radius for halos of $10^{12}$\Msun (blue), $10^{13}$\Msun
  (grey), and $10^{14}$\Msun (red).  Dark lines show the mean
  temperature or hot fraction at that radius for all selected halos,
  while shading shows the 1$\sigma$ scatter.  Dotted horizontal lines
  show the critical temperature we use to separate hot gas,
  $10^{5.4}$~K (top panel), and the critical hot fraction in our
  model, 0.6 (bottom panel).  $10^{12}$\Msun halos are hot out to
  $\sim 1 R_{\rm vir}$, while $10^{13}$\Msun and $10^{14}$\Msun halos
  are hot far beyond $R_{\rm vir}$, out to $3 R_{\rm vir}$.  The upper
  envelope of $10^{12}$\Msun halos remain hot out to several $R_{\rm
    vir}$, possibly because these halos are between $1-3 R_{\rm vir}$
  of more massive halos. }
\label{fig.fhot_vs_rad}
\end{figure}
\begin{figure*}
\includegraphics[width=160mm]{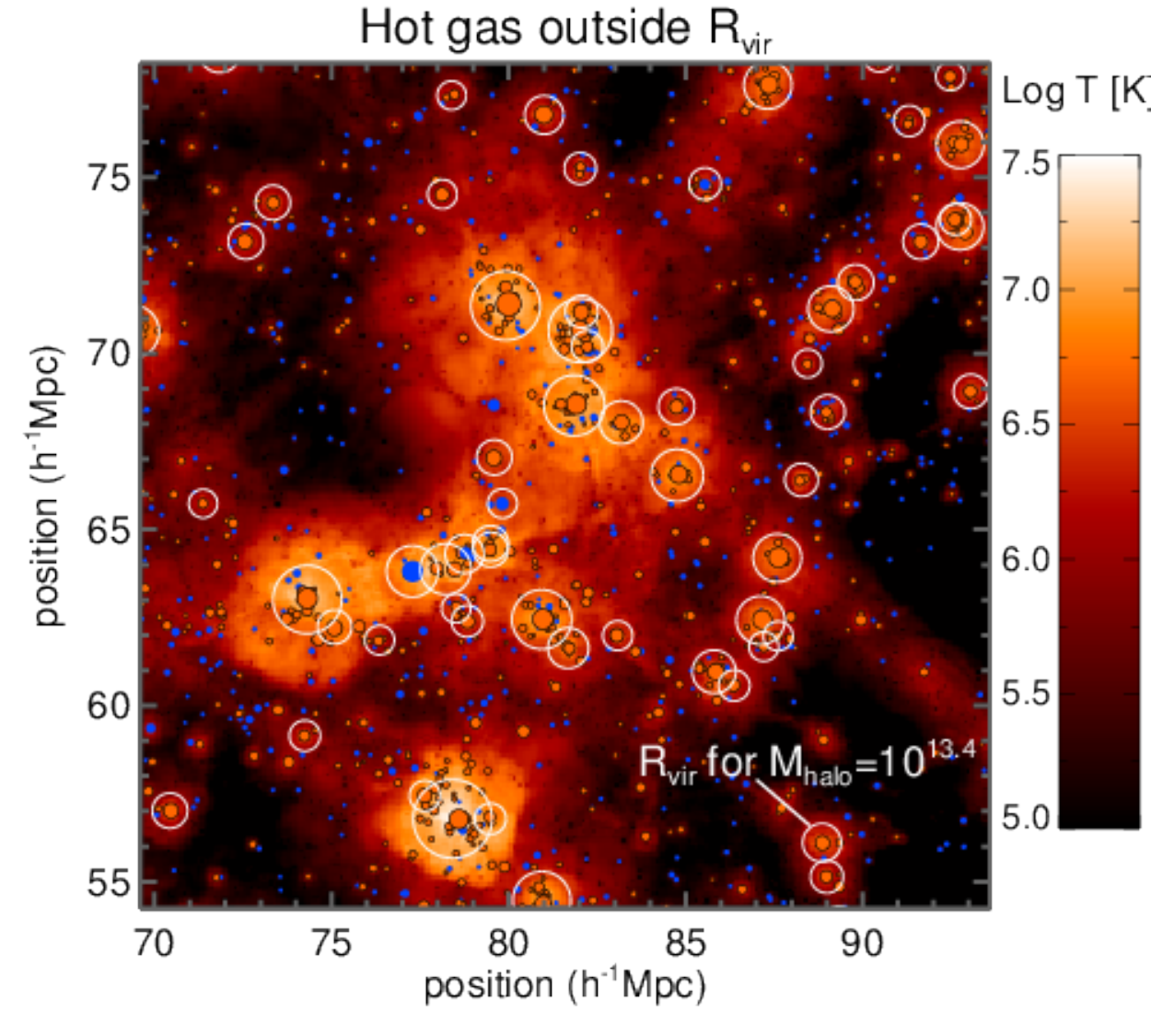}
\caption{This image centered on a super-cluster demonstrates hot gas
  outside halos in our simulation.  The background
  red-scale is the gas temperature.  On top we plot red and blue
  filled circles for red and blue galaxies, respectively, where the size of
  the circle scales as $\sqrt{M_{\rm stellar}}$.  White (unfilled) circles show
  the virial radii for halos with $M_{\rm halo} > 10^{13}$\Msun.  The
  most massive halo in this image ($M_{\rm halo}=2\times
  10^{14}$\Msun) appears near the bottom, just left of center. }
\label{fig.gasim_rvir}
\end{figure*}
An important implication of Figure \ref{fig.woo_dist_mhalo} is that
galaxies can be quenched well beyond the virial radius of a hot
gas dominated halo \citep[cf.][]{cybulski14}.  There are two effects that contribute to this:
First, there are ejected satellites, which we discuss further in
\S\ref{sec.lowmass_red}.  Here, we investigate how the hot gas
around massive halos can extend to well beyond the virial radius.

Figure \ref{fig.fhot_vs_rad} shows radial temperature profiles (top
panel), and the fraction of gas that is hot vs. radius (bottom panel)
for halos of mass $10^{12}, 10^{13},$ and $10^{14}$\Msun, out to $5
R_{\rm vir}$.  For $10^{12}$\Msun halos, the hot fraction normally
drops below the critical value of 0.6 just beyond the virial radius,
but for more massive halos the hot fraction remains high well beyond,
to $\approx 3 R_{\rm vir}$.  \citet{cen11} and \citet{bahe13} note similar effects in
their hydrodynamic simulations, in which the hot gas beyond the virial
radius also influences the gas content and star-formation of nearby
galaxies \citep[see also][who show that dark matter sub-halos
  falling into massive host halos lose mass well before they reach their
  hosts' virial radii]{behroozi13}.

We have explicitly checked that hot gas extends beyond the virial
  radius even in simulations with no hot gas quenching and in a
  simulation with neither quenching nor stellar-driven galactic winds.
  These simulations, which were run at the same resolution and in
  $48h^{-1}$~Mpc boxes, show hot fraction and temperature profiles
  very similar to those in Figure \ref{fig.fhot_vs_rad}.  Combined
  with the fact that such extended hot gas is noted by other authors
  using different hydrodynamic methods and feedback prescriptions,
  this result suggests that virial shocks, rather than sub-grid
  physics, create the extra-halo hot gas.  Hot gas beyond the virial
  radius is a general prediction of cosmological structure formation
  simulations.

We visually demonstrate hot gas beyond the virial radius in Figure
\ref{fig.gasim_rvir}.  Here we show a map of gas temperature only for
gas with temperature $T_{\rm gas}>10^{5.0}$~K, which appears as
red-scale in the figure.  We overplot blue and red galaxies in blue
and red circles, with the size of the circle scaling with the square
root of the stellar mass.  Finally, white circles show the virial
radii of halos with $M_{\rm halo} > 10^{13}$\Msun.  In the lower
right corner we indicate an example halo of mass $10^{13.4}$\Msun.
Hot gas clearly extends far beyond the virial radii of the most
massive halos.  It appears that the cosmic web forms super-structures
where hot gas pervades the regions between adjacent groups and clusters.

In summary, gas around massive groups and clusters ($M_{\rm
  halo}>10^{14}$\Msun) remains hot out to $\gtrsim 3 R_{\rm vir}$.
Hot gas pervades the regions between nearby groups and clusters, and
can lead to environmental quenching effects beyond the virial radii of
any massive halos.  We quantify the importance of this form of
quenching, which we call ``neighborhood quenching,'' in the next
section.

\subsection{Special case: Environment quenching of central galaxies, and ejected satellites}
\label{sec.lowmass_red}
\begin{figure}
\includegraphics[width=84mm]{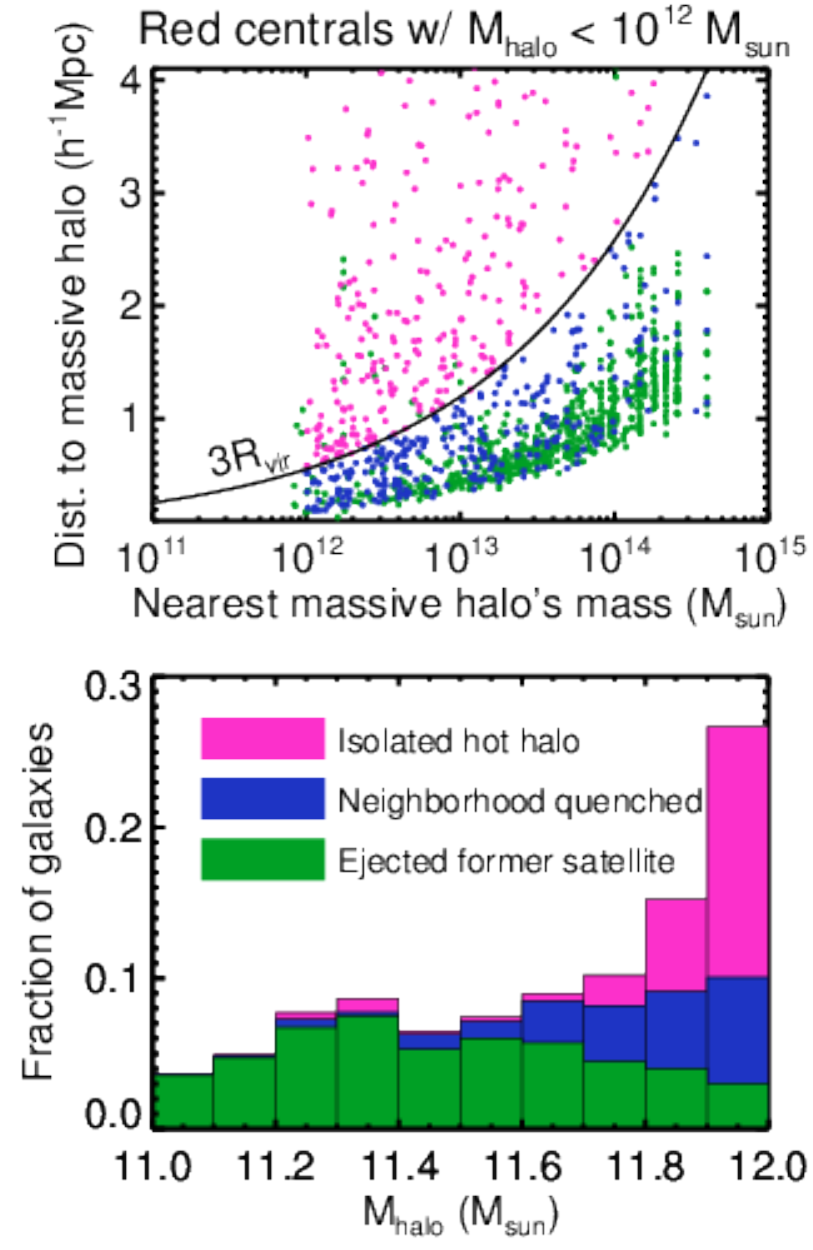}
\caption{{\bf Top:} Distance to the nearest massive halo vs. that
  halo's mass, for quenched central galaxies with halo masses
  $<10^{12}$\Msun at $z=0$.  These galaxies are quenched despite their
  low halo masses because (1) they are truly isolated halos with
  masses just under $10^{12}$\Msun, but they form hot halos owing to
  scatter in the hot gas fraction -- halo mass relation (pink points);
  (2) they are neighborhood quenched by hot gas that extends beyond
  the virial radius of massive halos (blue points); or (3) they are
  ejected former satellites of massive halos (green points).  We use
  the solid line marking $3 R_{\rm vir}$ to distinguish isolated hot
  halos (above the line) from neighborhood quenched galaxies.  {\bf
    Bottom:} Histogram of actual halo masses for red galaxies with
  $M_{\rm halo}<10^{12}$\Msun, with each bar in the histogram split by
  color depending on the galaxy's status: truly isolated, neighborhood
  quenched, or ejected from a massive halo.  Among red centrals with
  $M_{\rm halo}<10^{12}$\Msun, $\sim 50$\% are ejected satellites,
  $\sim 20$\% are neighborhood quenched, and $\sim 30$\% are isolated
  galaxies on the low-mass tail of the hot fraction vs. $M_{\rm halo}$
  relation (cf. Figure \ref{fig.hotfrac_vs_mass}).}
\label{fig.lowmass_red}
\end{figure}
Figure \ref{fig.mstellar_mhalo} revealed a population of quenched
central galaxies that live in halos with masses $<10^{12}$\Msun -- we
will call these low-mass red centrals.  In the standard ``halo
quenching'' picture where galaxies are quenched in halos
$>10^{12}$\Msun, these galaxies should not be quenched
\citep[cf.][]{cattaneo06}.  In our simulations, they are quenched for
(at least) three different reasons: (1) there is scatter in the hot
fraction -- $M_{\rm halo}$ relation as shown in Figure
\ref{fig.hotfrac_vs_mass}, so that some halos slightly below
$10^{12}$\Msun are quenching halos; (2) some galaxies are
``neighborhood quenched'': despite never having lived within the
virial radius of a hot halo, they are affected by the hot gas that
extends beyond the virial radius of massive halos (as shown in
\S\ref{sec.hot_beyond}, and discussed in \citealt{bahe13}); (3) some
galaxies are ejected former satellites of more massive halos, as
discussed at length in \citet{wetzel13}.


We disentangle these effects by identifying massive halos near the
low-mass red central galaxies.  For each low-mass red central galaxy
at $z=0$, we calculate a virial radius-scaled separation, $S_R \equiv
\Delta s / R_{\rm vir}$, from each halo with $M_{\rm
  massive}>10^{12}$\Msun.  Here $\Delta s$ is the separation between
the galaxy and the massive halo, and $R_{\rm vir}$ refers to the
virial radius of the massive halo.  We assign each galaxy (except
ejected satellites; see below) to the massive halo for which $S_R$ is
smallest, and refer to this halo as the nearest massive halo.


We also identify galaxies as ejected satellites by considering their
parent halo histories.  If a galaxy previously lived in a halo more
massive than its $z=0$ halo, and that previous halo had a mass
$>10^{12}$\Msun, then we consider it an ejected satellite.  In this
case, we find the most massive descendent of the previously
hosting halo, and assign that as the nearest massive halo.

In the top panel of Figure \ref{fig.lowmass_red}, we show the distance
to the nearest massive halo as a function of that halo's mass, for
these low-mass red centrals.  We show ejected satellites in green, and
split the remaining galaxies into neighborhood quenched (blue) and
truly isolated (pink).  To separate the latter two, we note from Figure
\ref{fig.fhot_vs_rad} that hot gas typically dominates out to $\sim 3
R_{\rm vir}$ for massive halos, and use this as a separator.  After
excluding ejected satellites, low-mass red centrals within $3 R_{\rm
  vir}$ of their nearest massive halos are designated as neighborhood
quenched, while those beyond $3 R_{\rm vir}$ are considered truly
isolated.

In the bottom panel of Figure \ref{fig.lowmass_red}, we show the
distribution of actual halo masses for low-mass red centrals.  Most of
these low-mass red centrals have halo masses just below
$10^{12}$\Msun.  We split each bar in the histogram by galaxy
category, as above.  About half of low-mass red centrals are ejected
satellites, $\sim 20$ per cent are neighborhood quenched, and $\sim
30$ per cent are isolated.  The distribution of isolated halos is
strongly peaked towards higher masses, indicating that these form the
low-mass tail of hot halos in the hot fraction-$M_{\rm halo}$
distribution (shown in Figure \ref{fig.hotfrac_vs_mass}).  That is,
these galaxies have formed their independent hot halos because they
have masses close to, but below, $10^{12}$\Msun.

To summarize, the population of quenched central galaxies with $M_{\rm
  halo} < 10^{12}$\Msun is roughly half ejected satellites, with a
significant contribution from galaxies that are ``neighborhood
quenched'' just because they live near massive, hot halos.
About 30 per cent are isolated halos with sufficient mass to support
their own hot coronae -- these form the low-mass tail of hot halos on
the hot gas fraction--$M_{\rm halo}$ relation (shown in Figure
\ref{fig.hotfrac_vs_mass}).

\section{Discussion}
\label{sec.discussion}

\subsection{A hot gas quenching narrative}
\label{sec.narrative}
Here we outline a quenching narrative that explains how galaxies are
quenched under different circumstances.  We consider galaxies whose
$z=0$ \emph{subhalo} masses end up at $10^{14}$\Msun, $10^{13}$\Msun,
$10^{12}$\Msun, and $10^{11}$\Msun.

First consider a massive central galaxy in one of the highest density
peaks of the cosmic web, destined to become the central galaxy in a
massive cluster ($M_{\rm halo}(z=0)>10^{14}$\Msun).  Such a galaxy rapidly
forms stars at high redshift, probably fueled by inflowing cold
streams of gas.  By $z\approx 2$ this massive galaxy's halo will
become dominated by hot gas.  The galaxy is strongly star-forming at
the highest redshifts, but transitions to red and dead around $z\sim
2$, depending on the halo mass \citep{dekel09, oser10}.  When its
star formation shuts down, this galaxy will live in the center of a
proto-cluster, and will remain quenched via some process that heats
the surrounding hot gas.  Thereafter, it will accrete additional 
stellar mass through mergers with satellite galaxies down to $z=0$.

Next, consider the galaxies in two \emph{subhalos} with final
masses of $10^{13}$\Msun: galaxy A in a highly
overdense region near a cluster, galaxy B isolated.  At high
redshifts, both galaxies rapidly form stars, then their halos grow
sufficiently large to support a hot halo and quenching begins around
$z\sim 1-2$.  This path is similar to that for the central cluster
galaxy described above, but the quenching occurs later.  After being
quenched, galaxy A will eventually join a massive cluster, becoming a
massive satellite within the cluster.  It may grow in mass by
continuing to accrete its own satellites even after joining the
cluster.  Galaxy B remains isolated, becoming the central group galaxy
of its own group.  Both of these galaxies are mass quenched, but
galaxy A ends up as a satellite.

Galaxies with final subhalo masses of $\sim 10^{12}$\Msun have at
least 5 different possible evolutionary histories.  Galaxy I is in a
relatively underdense environment.  It evolves as a star-forming
galaxy for most of cosmic history, then its halo becomes hot at low
redshift and it appears as an isolated quenched galaxy at $z=0$.
Galaxy J is in a similarly underdense environment, but its halo
remains dominated by cold gas until $z=0$, simply
because of the scatter in the hot gas fraction--halo mass relation
(Figure \ref{fig.hotfrac_vs_mass}) -- there are halos slightly above
$10^{12}$\Msun that are never dominated by hot gas.  Galaxy J thus
remains as a star-forming galaxy throughout cosmic history.  Galaxy K,
on the other hand, lives in an overdense region.  It evolves mostly
like galaxy I, except that after quenching at low-redshift, it joins a
more massive halo and becomes a satellite galaxy.  It is pre-processed
in its own halo.  Galaxy L joins a cluster before it is quenched by
its own halo, so it is satellite quenched and remains at a lower
stellar mass than galaxy K.  Galaxy M joins a group before being being
quenched by its own halo -- it is satellite quenched in the group --
but then the host group merges with a cluster.  Galaxy M is thus
pre-processed as a satellite galaxy in a group before ending up in a
cluster.

Finally, consider galaxies with final subhalo masses of $\sim
10^{11}$\Msun.  These generally take similar paths as the
$10^{12}$\Msun galaxies above, except that they are never quenched in
isolation by their own halos.  A small number may, however, be
quenched in a massive group or cluster then be ejected from that halo,
thus appearing as low-mass quenched centrals
(cf. \S\ref{sec.lowmass_red}).  It is also possible that they are
``neighborhood quenched'' simply by living \emph{near} a massive hot halo
whose hot gas extends well beyond its own virial radius.

These narratives illustrate the wide diversity of quenching histories.
All are seen to occur in our simulations, arising from a simple criterion
of preventing cold gas infall whenever the halo hot gas fraction exceeds 
60\%.  Since so many different narratives occuring in a wide range
of environments can lead to quenching of galaxies around $10^{12} M_\odot$, 
the natural consequence is that mass quenching
is mostly independent of environment.  Meanwhile, satellites that fall
into or live near large halos are quenched, irrespective of their mass,
which leads to the conclusion that environment (or more appropriately
satellite) quenching is roughly independent of mass.

\subsection{Galaxy structure in this quenching narrative}

The narrative laid out above makes no mention of galaxy structure,
morphology, or kinematics, despite that observations link these properties
to quenching.  Since our simulations cannot resolve thin
stellar disks, they do not directly address the relationship between
quenching and galaxy structure.  

Quenched galaxies are generally bulge-dominated in the local universe,
and recent work links central stellar density
(\citealt{kauffmann03_structure}, \citealt{fang13} at low-$z$; and
\citealt{franx08}, \citealt{cheung12} at higher-$z$), stellar velocity
dispersion \citep{bezanson13}, or bulge mass \citep{bluck14} to
quenching.  This may suggest an important role for gas-rich major
mergers, which commonly form classical bulges in idealized simulations
\citep[e.g.][]{barnes90, hernquist93, robertson06_fundy_plane}.  But a
substantial population of disk galaxies lives on the red sequence in
both the local universe \citep{skibba09_zoo, masters10} and at $z>1$
\citep{bundy10, vanderwel11, bruce13}, and most quenched ellipticals
show ordered rotation
(\citealt{emsellem07}; \citealt{krajnovic11}; \citealt{emsellem11};
though \citealt{cox06_kinematics} show that merger remnants may show ordered
rotation).

So how could galaxy structure play into our hot gas quenching
narrative?  One intriguing possibility is that massive galaxies
establish dense stellar cores at early times, before quenching.  Halos
in the high-redshift universe are thought to be denser than those
today, and high-redshift ($z\gtrsim2$) star-forming galaxies may
undergo disk instabilities \citep{noguchi99, bournaud07_clumps,
  dekel09, ceverino10} that help build observed dense star-forming
bulges \citep{barro13a, barro13b, williams14}.  Perhaps after a
starburst that exhausts their remaining dense gas, they quench in hot
halos that starve them of fuel for star-formation.  As their remaining
disks fade and are harrassed by close encounters in dense environments, 
they become strongly bulge-dominated in appearance
\citep{carollo13}.  After this, they may grow in stellar mass and size
through mergers \citep{naab09, oser12, gabor12}, with the details of
their merger histories determining their final kinematics and internal
structure \citep{naab13}.  This scenario is essentially the one
advocated by \citet{dekel13_nuggets}.  Our simulations imply that
later gas accretion must be prevented for such galaxies to remain red
\citep{gabor11}; hot gas quenching is one such prevention
mechanism that naturally leads to this scenario.


Gaseous mergers may also play a role in quenching, even if they are
not solely responsible for it.  Once a galaxy is starved of new fuel,
it must rely on consuming its existing fuel to remain blue.  If during
this vulnerable stage it undergoes a merger, this can result in a
rapid consumption of remaining gas (including the ejection of
substantial cold gas in a molecular outflow; \citealt{walter02,
  narayanan06, narayanan08, feruglio10, sturm11}), thereby
accelerating the transition to the red sequence.  Hence in the hot gas
quenching scenario, the role of gaseous mergers may be to set the
\emph{rapidity} of the transition to the red sequence.  The most
rapidly transitioning systems would appear as post-starburst
galaxies~\citep{quintero04, yang08, wild09, snyder11}, but galaxies
that do not undergo such a merger would live as red disks for some
time before their disk morphologies fade.  Observations can constrain
this dichotomy in quenching timescales~\citep{schawinski14}, but simulating such
morphological transitions together with quenching directly will
require very high resolution in large volumes with representative
galaxy samples -- a challenge given current computational
capabilities.

\section{Conclusion} 
\label{sec.conclusion} 
Using cosmological hydrodynamic simulations with a simple model for
quenching star formation in massive halos dominated by hot gas, we
have shown that hot gas drives both mass quenching and environment quenching.
The model reproduces the crucial $z=0$ trends among
quenching, stellar mass, halo mass, environment, and distance to the
halo center.  It also provides physical explanations for the observed trends.
Key results are that:
\begin{itemize}
\item Our model reproduces the observed trend among red fraction,
  environment, and stellar mass that leads to the interpretation of
  separable mass quenching and environment quenching (Figure
  \ref{fig.2dhists}).  Galaxies are increasingly likely to be quenched
  at high stellar masses and in dense environments.  This trend
  emerges naturally from our hot gas fraction cut, or even a halo mass
  cut: galaxies are increasingly likely to be hosted by a halo with
  $M_{\rm halo}>10^{12}$\Msun at high stellar masses and in dense
  environments.  
\item We also explicitly show that mass quenching mainly
  applies to central galaxies and environment quenching applies to
  satellite galaxies (Figure \ref{fig.fsat}).
\item The stellar mass -- halo mass relation in our model matches
  observationally constrained abundance matching in the local universe
  (Figure \ref{fig.mstellar_mhalo}), a key barometer that is difficult
  to achieve in current hydrodynamic simulations.
\item Mass quenching appears independent of environment because halos
  with $M_{\rm halo}\sim 10^{12}$\Msun, where quenching begins, exist
  in a wide range of environments (Figures \ref{fig.2dhists} and
  \ref{fig.env_mhalo}).  This includes isolated ``field'' environments
  (Figure \ref{fig.gas_im}).
\item Environment quenching appears independent of stellar and halo
  mass because galaxies in the densest environments uniformly live in
  hot halos with $M_{\rm halo}>10^{12}$\Msun.  Once a galaxy lives in
  such a hot halo, it will be quenched regardless of its stellar mass
  or host halo mass.
\item Satellite galaxies show the effects of mass quenching,
  independent of their host halo mass (Figure \ref{fig.bluesat_mf}).
  This is because massive galaxies that end up as satellites at $z=0$
  are typically mass quenched in their own hot halos at higher
  redshift -- a kind of pre-processing (cf. Figure
  \ref{fig.halo_hist}).
\item About $1/3$ of red satellites were pre-processed, including
  $\sim 90$ per cent of red satellites with $M_{\rm
  stellar}>10^{10.5}$\Msun.  About $2/3$ of these massive red
  satellites were central galaxies when first quenched: they were mass
  quenched in their own hot halos, then joined larger halos to become
  satellites (Figure \ref{fig.fprocess}).
\item Satellite galaxies are increasingly likely to be quenched at
  higher host halo masses and closer to the center of their host halos
  (Figure \ref{fig.woo_dist_mhalo}).  This trend matches observations.
\item Hot gas can extend far beyond the virial radii of massive halos,
  moreso for larger halos, reaching $\sim3~R_{\rm vir}$ for clusters 
  (Figures \ref{fig.fhot_vs_rad} and \ref{fig.gasim_rvir}).
\item Our model produces a population of red central galaxies that
  live in low-mass halos ($M_{\rm halo}<10^{12}$\Msun).  These form
  through three main mechanisms: over half are ejected satellite
  galaxies that once lived in a more massive group or cluster hot
  halo; about 20 per cent are ``neighborhood quenched'', embedded in the hot
  gas beyond the virial radius of a neighboring group or cluster; and
  about 30 per cent are truly isolated, arising from the scatter in the hot
  fraction -- halo mass relation (Figure \ref{fig.lowmass_red}).  
\end{itemize}

Our model outlines a plausible and simple scenario for the formation of 
quenched central and satellite galaxies in the nearby universe.  It
does not directly account for the morphology of galaxies, but including
some simple considerations of morphological transformation owing to
mergers, it seems at least qualitatively to accommodate a range of
quenching paths from rapid quenching through a merger-induced post-starburst
phase to a more gentle quenching via a red disk phase; in essence, the
rapidity of quenching is set by the random chance of suffering a merger
event, but the quenching itself is driven purely by the existence
of a hot halo.

This heuristic model leaves many unanswered questions, of course.  The
main one is, what keeps the hot halo gas hot?  The putative answer is
AGN feedback, although the driving physics remains to be fully worked
out.  Another important consideration is the effect of galactic
outflows during the star-forming phase that pumps metal-enriched (and
thus more rapidly cooling) gas into the massive halo's circumgalactic
medium.  Such outflows could potentially either add heat or enhance
cooling~\citep{keres09_feedback}, and could interact with inflowing
filaments in ways that are difficult to model a
priori \citep{keres09_coldclouds} and difficult to understand
analytically.  There are also important redshift-dependent effects
that hot halos may be less effective at preventing accretion at
earlier epochs~\citep{dekel09}.  Hence there remains much work to be
done to sort out the physical drivers behind and cosmic evolution of
hot gas quenching.

 
 

\section*{Acknowledgements}
We thank Joanna Woo for the observational data for Figure 11.  We
acknowledge Volker Springel for making \textsc{Gadget} publicly
available, and thank the referee for helpful comments. JMG
acknowledges support from the EC through grants ERC-StG-257720 and the
CosmoComp ITN.  Simulations were performed at TGCC and as part of a
GENCI project (grants 2011-042192 and 2012-042192).

\bibliographystyle{mn2e} 

\bibliography{paper}


\label{lastpage}

\end{document}